%% file: ms.tex
\documentclass{article}
\usepackage[utf8]{inputenc}
\usepackage[titletoc,toc,page]{appendix}
\usepackage{authblk}  
\usepackage{url}
\usepackage{float}
\usepackage{amsthm}
\usepackage{amsmath}
\usepackage{esint}
\usepackage{amssymb}
\usepackage{geometry}
\usepackage{graphicx}
\usepackage{subcaption}
\usepackage{adjustbox}
\usepackage{mwe}
\usepackage{tikz}
\usepackage{eso-pic}
\usetikzlibrary{shapes.geometric}
\usetikzlibrary{hobby}
\usetikzlibrary{fadings}
\tikzfading[name=fadeIn, inner color=transparent!100,outer color=transparent!20]
\usepackage{pgfplots}
\usepackage[bottom]{footmisc}
\usetikzlibrary{decorations.shapes}
\pgfplotsset{compat=newest}
\ifpdf
    \usepackage{forest}
    \usetikzlibrary{backgrounds}
\else
    \usepackage{pst-barcode}
\fi
\colorlet{dens0}{red}
\colorlet{dens1}{purple}
\colorlet{dens2}{green}
\colorlet{dens3}{brown}
\colorlet{dens4}{blue}
\colorlet{rBlue3}{red!25!blue}
\colorlet{rBlue1}{red!75!blue}
\colorlet{rBlue2}{red!50!blue}
\colorlet{red1}{red!75!white}
\def\width{18}
\def\hauteur{9}
\tikzset{paint/.style={ draw=#1!50!black, fill=#1 },
    decorate with/.style=
    {decorate,decoration={shape backgrounds,shape=#1,shape size=1mm}}}
\tikzset{
    graphnode/.style={
      draw,fill,
      circle,
      minimum size=0.7mm,
      inner sep=0
    },
}
\newcommand{\NFM}[1]{\adjustbox{scale={0.8}{1.5}}{\bf\scriptsize #1}}
\newtheorem{theorem}{Observation}
\newtheorem{dfn}{Definition}
\title{\LARGE \bf
The r\^ole of city geometry in determining the utility of a small urban light rail/tram system
}

\author{Michael Mc Gettrick\thanks{michael.mcgettrick@nuigalway.ie}}
\affil{School of Mathematics, Statistics and Applied Mathematics, National University of Ireland, Galway, Ireland}

\begin{document}

\maketitle
\tableofcontents


\begin{abstract}

In this work, we show the importance of considering a city's \emph{shape}, as much as its population density figures, in urban transport planning.
We consider in particular cities that are ``circular'' (the most common shape) compared to those that are
``rectangular''.
For the latter case we show greater utility for a single line light rail/tram system.
A particular case study is presented for Galway City.
\end{abstract}

\section{Introduction}

There are many factors to consider when constructing a light rail / tram system in a city. Some of these factors can be scientifically analyzed (as is the case in this paper), but others perhaps not (aspects that are political, sociological, financial,....). These latter aspects are of course important, but are not analyzed here.

Further, amongst considerations that lend themselves to scientific analysis, we restrict further to 
\emph{small} tram systems, indeed while we look a little at a two line system, most of this paper considers a single line system. To make the model tractable, in this current work we restrict ourselves to circular or rectangular ``geometries'' (which should none the less approximate many real world city shapes).

\section{City Geometries}

\subsection{Rectangular vs. circular}\label{sec:Rect}
We analyze here idealized models where the 
city shape is either rectangular (of dimensions $p$ (kilometres) times $lp$, where $p$ is the short side of 
the rectangle and $l\geq 1$) or circular (of radius $r$ kilometres).
For the rectangle, $p$ defines its ``size'', while $l$ defines its ``shape'' (varying from $l=1$ (square)
to large values of $l$ for a ``long'', ``skinny'' rectangle).
As a (very) simplifying assumption, we assume uniform population density inside the city shape - the results of the analysis in this paper only apply in citys that approximate this distribution.\footnote{So, the results here will not apply in ``large'' cities, where there are skyscrapers / tower blocks / large apartment blocks in city centres. They will also not apply in small cities in countries where there is a tradition of people living in apartment blocks in city centres (much of continental Europe, for example). But our results will apply in smaller cities in USA, UK, Ireland, for example, which generally do not have large apartment blocks in their centres.}

Assuming a uniform population density inside the city, let us suppose a person wants to go from a random location $i$ to a random location $j$ inside the city. One question we may ask is, what is the average length $\Bar{d}$ of such trips? It should be clear that this average distance will be an increasing function of all our
parameters $r, p, l$. It should also be intuitively clear that,
given a ``rectangular'' city and a ``circular'' one of equal
area (and the aforementioned uniform population density), if
$l\gg 1$ then $\Bar{d}$ will be greater in the rectangular case.
These mathematical questions are studied in the discipline of Geometric Probability, see
\cite{gp2, gp1}.

\begin{figure}[ht!]
\centering
\includegraphics[width=12cm, trim={0 0.5cm 0 0},clip]{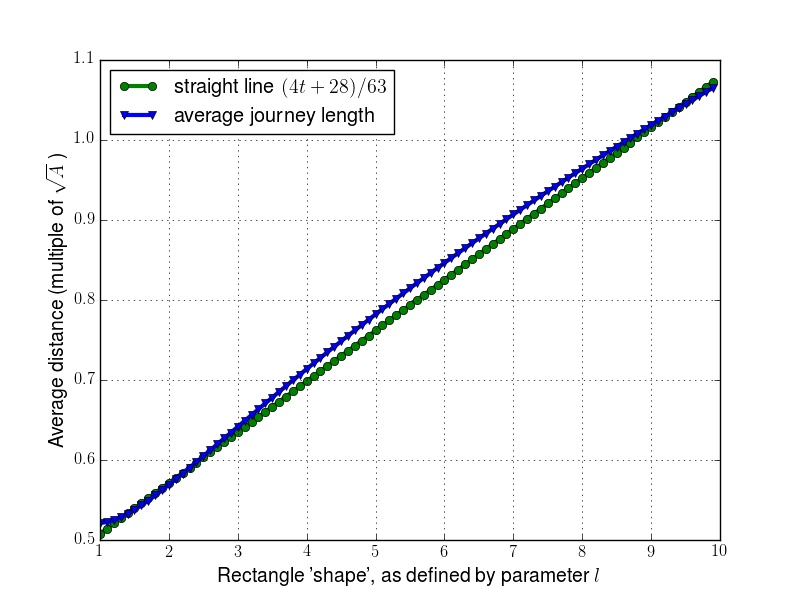}
\caption{\small The average distance ($\Bar{d}^{\rm rect}$)
between two 
randomly chosen points in a rectangle of dimensions
$(p)\times (lp)$, and fixed area $A=lp^2$. The curve in blue (triangles) is a plot of equation (\ref{eqn1}). The intercept on the y-axis ($l=1$) corresponds to a square (see equation (\ref{eqn3})).
\label{fig:avDist}}
\end{figure}

The analysis of \cite{circle1} gives for the circular case
\begin{equation}\label{eqn2}
\Bar{d}^{\rm circ} = \frac{128r}{45\pi},
\end{equation}
while for the rectangular case \cite{Mathai} gives
\begin{equation}
\Bar{d}^{\rm rect} = \frac{p}{15}\left[l^3 +\frac{1}{l^2} + \sqrt{l^2+1}\left(3-l^2-\frac{1}{l^2}\right)
+ \frac{5}{2l}\ln \left(l+\sqrt{l^2+1}\right) + \frac{5l^2}{2}\ln \left(\frac{1+\sqrt{l^2+1}}{l}\right)\right].
\end{equation}
A natural question to ask is, how does $\Bar{d}^{\rm rect}$ vary, for a city of fixed area, as the shape varies from
square to a long/thin rectangle? For fixed $A=lp^2$, the $l$ dependence is given by
\begin{equation}\label{eqn1}
\Bar{d}^{\rm rect} = \frac{\sqrt{A}}{15\sqrt{l}}\left[l^3 +\frac{1}{l^2} + \sqrt{l^2+1}\left(3-l^2-\frac{1}{l^2}\right)
+ \frac{5}{2l}\ln \left(l+\sqrt{l^2+1}\right) + \frac{5l^2}{2}\ln \left(\frac{1+\sqrt{l^2+1}}{l}\right)\right]
\end{equation}
which presents as a curve with a reasonably uniform slope close to $\sqrt{A}/15$ (see figure \ref{fig:avDist}).
Note further that for the case of a 
square ($l=1$), equation (\ref{eqn1}) gives 
\begin{equation}\label{eqn3}
\Bar{d}^{\rm square} = \left(\frac{2+\sqrt{2}+5\ln(1+\sqrt{2})}{15}\right)p
\approx 0.52p
\end{equation}
which is slightly larger than (\ref{eqn2}) (setting $p=2r$), as expected.
\begin{theorem}
For rectangular cities, the average distance travelled by inhabitants is larger the more rectangular the city is.
Therefore, the need for a rail/tram system is larger.
\end{theorem}

\subsection{Where to put the rail/tram line?}
In what follows, we will just use the word ``tram'', instead of mentioning rail/tram each time.

\subsubsection{Rectangular city}
Suppose we put a single line tram in our ``rectangular'' city. Assume that residents will consider using
the tram if they live within $d_t$ kilometres of it (later we will set $d_t$ to 0.5km). In our idealized model,
we put the tram along the centre of the rectangle, running along the longer side, stopping at each end $d_t$
from the end. To explain it using the cartesian plane, if our rectangular city is placed with the shorter side
going from $(0,0)$ to $(0,p)$, and the longer side from $(0,0)$ to $(lp,0)$, then the tram line will run 
from $(d_t,p/2)$ to $(lp-d_t,p/2)$. We assume the tram has $n$ stops along the line (including the ends), and 
an average of $\tau$ minutes to go between stops (so one full run along the line takes $\tau (n-1)$ minutes).

The average speed of the journey between any two tram stations for the rectangular city is simply
\begin{equation}
\Bar{s}^{\rm rect} = \frac{lp-2d_t}{(n-1)\tau}
\end{equation}

\subsubsection{Circular city}
In what follows, we assume a circular city of equal area to the rectangular one (and of equal population,
hence of equal population density). We have that $lp^2 = A = \pi r^2$, where $r$ is the radius (in kilometres)
of the city. Since we assumed $l \geq 1$, $p$ must be less than the diameter of the circle ($p < 2r$).
\begin{description}
\item[1 line]
Starting at $l=1$, our rectangular city is a square.
Since $lp^2 = A = \pi r^2 = p^2$, we have that $p=r\sqrt{\pi} \approx 1.77 r$, so the sides of the square are
less than the circle diameter. As we increase $l$, keeping areas constant, our line of length 
$lp-2d_t$ becomes longer until its length equals that of the longest straight line (tram) we put in the 
circle, of length $2r-2d_t$, so we get 
\begin{equation}\label{smallL}
    lp = 2r = 2p\sqrt{\frac{l}{\pi}} \implies \sqrt{l} = \frac{2}{\sqrt{\pi}} \implies 
    l = \frac{4}{\pi} \approx 1.27
\end{equation}
\item[2 lines]
Increasing $l$ further above $4/\pi$, we match this (one) line in the rectangle with two intersecting lines 
in the circle, whose  lengths sum to  $lp-2d_t$. In our idealized model, we run the two lines at right angles
to one another, intersecting at the centre of the circle. As $l$ increases, eventually the sum of
the maximum lengths of 
the two lines in the circle ($4r-4d_t$) must match $lp-2d_t$, so we get
\begin{equation}\label{bigL}
    4r-2d_t = lp = lr\sqrt{\frac{\pi}{l}} 
    =r\sqrt{\pi l}
    \implies
    \sqrt{l} = \frac{2}{\sqrt{\pi}} \left( 2 - \frac{d_t}{r}\right)
    \implies l \approx
    \frac{16}{\pi} \left( 1 - \frac{d_t}{r} \right)
\end{equation}
where we ignore the terms quadratic in $d_t/r$.
We anticipate (later) that $d_t/r$ will be approximately 0.1, and we may ignore this in 
later approximations.

\end{description}
\begin{theorem}
Obviously the circular city with 2 lines also demands other infrastructure - an intersection tram
station at the center, where the two lines meet, to allow passengers to swap lines. This junction may 
imply further cost and/or restrictions:
\begin{enumerate}
    \item {\bf Cost:} Build a full overpass/underpass system (bridge) so trams pass freely.
    \item {\bf Restriction:} In the absence of a bridge, there may be a restriction on schedules (2 trams
    can't pass the junction at the same time), perhaps causing inferior service.
\end{enumerate}
\end{theorem}

\section{``Quality of Service''}
While we cannot ``guesstimate'' scientifically how many people may use a tram service, we 
can say that the better the quality of service, the more will use it.
Two main factors (amongst others) that affect quality of service are
\begin{enumerate}
    \item 
    Frequency of service
    \item
    Time to get ``from A to B''
\end{enumerate}
These considerations may not be independent: In the rectangluar model, the travel time does not 
depend on the service frequency, but in the circular model, it may, in trips where one has to 
transfer lines.
In this section we analyze the average time required to get ``from A to B'' in our ``circular'' and 
``rectangular'' cities. Let $\lambda$ be the \emph{average} distance between neighboring stops on the line,
so, the \emph{average} speed is 
\begin{equation}\label{avSpeed}
\Bar{s} = \lambda / \tau
\end{equation}
If there is a tram every $t_f$ minutes (frequency of service), the 
average waiting time for a tram is $t_f/2$: It will be convenient in what follows to write this time 
in terms of $\tau$,
\begin{equation}\label{tau1}
t_f/2 = q\tau
\end{equation}
and we will consider situations where $q=1, 2, 3,\dots$.

We will call the \emph{region of service} of the tram the set of all points $(x,y)$ that are within distance $d_t$ of the tram line. Within this region of service, we now compare the average speed of a journey between two tram stations, for the case defined by equation (\ref{bigL}), for our circular and rectangular cities.
(Note that for the case of equation (\ref{smallL}), this average speed will be identical in both cases, and equal to $\lambda/\tau$.)

\begin{figure}[ht!]
\centering
\includegraphics[width=50mm]{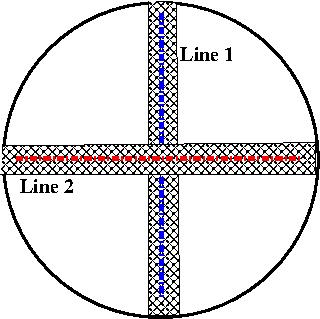}
\caption{\small Two tram lines (red and blue) intersecting at the centre: the shaded region is the region of service. \label{twoLines}}
\end{figure}

\begin{figure}[ht!]
\centering
\includegraphics[width=17cm, trim={0 15.5cm 0 0},clip]{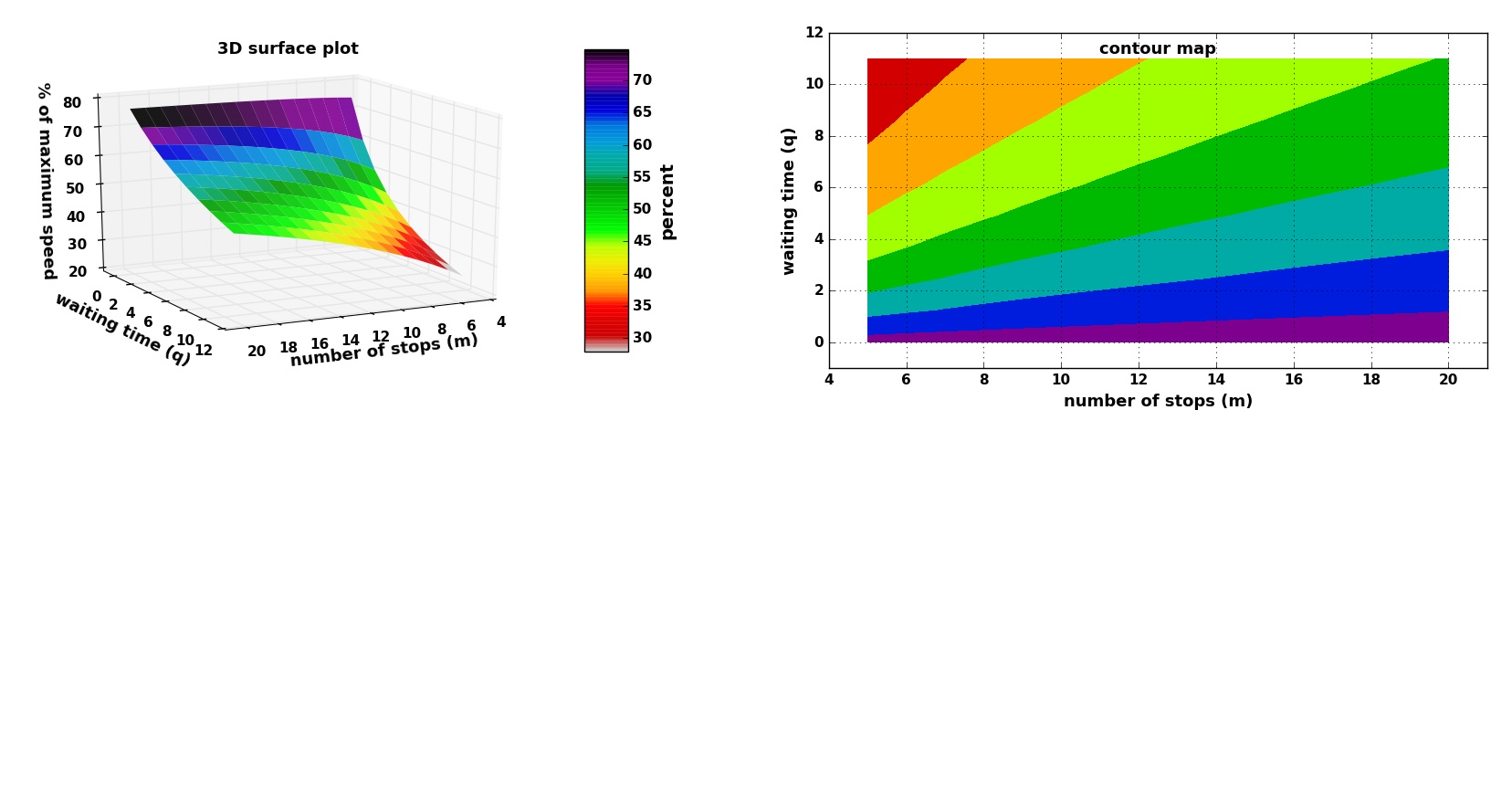}
\caption{\small The average speed between randomly chosen 
stations on \emph{different} tram lines (see figure
\ref{twoLines}). All parameters are dimensionless. The average speed is written as a percentage of $\Bar{s} = 
\lambda / \tau$ (see equation 
\ref{avSpeed}). $m$ is the total number of stations (counting from the origin / junction point of the two lines) while $q$ is the average waiting time at the junction as a multiple of $\tau$ (see equation \ref{tau1}).\label{fig:2Lines}}
\end{figure}

\begin{description}
\item[ Rectangular city: ]
The average speed is $\lambda/\tau$.
\item[ Circular city: ]
For journeys using only one single line (i.e.\ 
from stations on the blue line to other stations on 
the blue line, or from stations on the red line to 
other stations on the red line), the average speed is
$\lambda/\tau$ as before.
However, if we make a journey between two randomly chosen stations, half of these journeys involve changing line, and we will show this results in 
considerably reduced speed.

Letting the origin be at the usual position of the intersection of the $x$ (red) and $y$ (blue) lines, we label the stations with integers $i$ (along $x$) and $j$ (along $y$), so that the average distance between
station $i$ and $j$ is $\lambda\sqrt{i^2 + j^2}$.
There is a certain amount of waiting time at the junction at $(0,0)$, which we write in terms of 
$\tau$ as described in equation [\ref{tau1}]. Then
the total transit time from $i$ to $j$ is 
$(i+j+q)\tau$. This results in an average speed from $i$ to $j$ of 
\begin{equation}
    s_{ij} = \frac{\lambda\sqrt{i^2 + j^2}}{(i+j+q)\tau}
\end{equation}
Since the numerator is less than $(i+j)\lambda$, and
the denominator is greater than $(i+j)\tau$, this speed can be considerably less than $\lambda /\tau$. Averaging over all $i, j \leq m$ gives an average speed
of
\begin{equation}
    \Bar{s} = \frac{1}{m^2}\left(\frac{\lambda}{\tau}\right)\sum_{i,j = 1}^m \frac{\sqrt{i^2 + j^2}}{(i+j+q)}
\end{equation}
We plot $\Bar{s}$ as a function of $m$ and $q$ in figure \ref{fig:2Lines}.

\begin{theorem}
Note that for a substantial range of the parameters $m$ and $q$, the average speed falls to $50\%$.
\end{theorem}
\begin{theorem}
For fixed $m$, the larger the value of $q$ (i.e. the larger the wait at the junction), the smaller the overall point-to-point speed.
\end{theorem}
\begin{theorem}
For fixed values of $q$, as we increase $m$ we reduce the impact of $q$ on the overall speed.
\end{theorem}
\end{description}

.

\section{Infeasible Regions}
For all choices of departure point $(x_1, y_1)$ and arrival point $(x_2, y_2)$, we now compare travelling with or without the tram. If we fix a particular departure point $(x_1, y_1)$ and consider all possible arrival points 
$(x_2, y_2)$, it is intuitively clear that for some arrival points, using the tram would serve no purpose (we say the
journey is infeasible via the tram). In this section we define this notion precisely, and calculate what proportion
of trips are infeasible. We show that the infeasible region for a circular city is larger than that for a rectangular city.
\subsection{Metrics}
We denote by $d_{\rm euc}$ the standard Euclidean metric on
the plane $\mathbb{R}^2$, with the distance between two
points $p_1, p_2 \in \mathbb{R}^2$ given by 
\begin{equation}
    d_{\rm euc}(p_1, p_2) = \sqrt{(x_2-x_1)^2 + (y_2-y_1)^2}
\end{equation}
where $p_1 = (x_1, y_1)$ and $p_2 = (x_2, y_2)$.

We now define a metric for measuring the (effective) distance  for the journey between $p_1$ and $p_2$ using the
tram (see \cite{metric1} for general information
on metrics). This is comprised of three parts:
\begin{enumerate}
    \item The journey (not using tram) from $p_1$ to the nearest tram station
    \item The journey on the tram to the station nearest the destination
    \item The journey (not using tram) from that station to the destination ($p_2$)
\end{enumerate}
To calculate this we need to determine the nearest point on the tram line to an arbitrary point $p = (x,y)$. Let our
tram line run between points $(-a,0)$ and $(a,0)$, so it
is represented by a horizontal line segment centered at the origin $(0,0)$. Let us define
\begin{equation}
    x^t = 
    \begin{cases} 
-a, & x < -a\\
x, & -a \leq x \leq a\\ 
a, & x > a\\
\end{cases}
\end{equation}
or equivalently
\begin{equation}
    x^t = 
    \begin{cases} 
\max (x,-a), & x < 0\\
\min (x,a), & x > 0\\ 
\end{cases}
\end{equation}
The nearest point on the tram line to any point
$p = (x,y)$ is $Proj(p) = (x^t,0)$. Our tram metric $d_{\rm tr}$ is
\begin{equation}
    d_{\rm tr} (p_1,p_2) = 
    d_{\rm euc} (p_1, Proj(p_1))
    + |x_1^t - x_2^t|(\Bar{s}_{\rm nt}/\Bar{s}_{\rm t})
    + d_{\rm euc} (Proj(p_2), p_2)
\end{equation}
where $\Bar{s}_{\rm t}$ is the average speed along
the tram line, while 
$\Bar{s}_{\rm nt}$ is the average
speed on the other two (non-tram) segments of the journey. Note here that the physical distance 
$|x_1^t - x_2^t|$ along the tram line
is reduced by a factor of $\Bar{s}_{\rm nt}/\Bar{s}_{\rm t}$
because of the superior speed of the tram $\Bar{s}_{\rm t}$
(compared to 
$\Bar{s}_{\rm nt}$)\footnote{For example: By foot, 
$\Bar{s}_{\rm nt} \approx 0.1$ km/minute, by bicycle
$\Bar{s}_{\rm nt} \approx 0.2$ km/minute. On the Paris
metro, $\Bar{s}_{\rm t} \approx 0.5$ (measured by the author on line 4, between Jussieu and Mairie d'Ivry, November 2018), while the 
Dublin Luas has $\Bar{s}_{\rm t} \approx 0.28$}.

\subsection{Infeasible Journeys}
To compare $d_{\rm euc} (p_1, p_2)$ 
with
$d_{\rm tr} (p_1, p_2)$
we have to consider one further term: Travelling via
the tram, when one arrives at the point $Proj(p_1)$
one must wait on average time $t_f/2$ for a tram. In 
this time, \emph{if} the traveller had not taken the 
tram, they could have travelled a distance of 
$\Bar{s}_{\rm nt}t_f/2$ kilometers. Thus, we should 
compare $d_{\rm euc} (p_1, p_2)$ with
$d_{\rm tr} (p_1, p_2)  + \Bar{s}_{\rm nt}t_f/2$.

\begin{dfn}
We say a journey between points $p_1$ and $p_2$ is
\emph{($\alpha$,$\beta$)--infeasible} if and only if
\begin{equation}\label{infeas2}
    d_{\rm euc} (p_1, p_2) - d_{\rm tr} (p_1, p_2)
    < \Bar{s}_{\rm nt}t_f/2
\end{equation}
where $\alpha = \Bar{s}_{\rm t} / \Bar{s}_{\rm nt}$ is 
the tram speed relative to the non-tram speed, and
$\beta = t_f/2$ is the average waiting time.
\end{dfn}

\begin{dfn}\label{defTwo}
We say a journey is \emph{infeasible} if and only if it is
\emph{($\infty$,$0$)}--infeasible.
\end{dfn}
Definition (\ref{defTwo}) corresponds to an ideal
world of zero waiting time for the tram, and 
infinite tram speed!

\begin{dfn}
The \emph{infeasible region} corresponding to a point
$p$ is the set of all points $q$ such that the journey
from $p$ to $q$ is an infeasible journey.
We denote this using the function 
\begin{equation}
I: \mathbb{R}^2 \to 
2^{\mathbb{R}^2}, \ 
I(p) = \{ q \in \mathbb{R}^2 | 
d_{\rm euc} (p,q) < d_{\rm tr} (p,q) \}
\end{equation}
\end{dfn}

For a city whose area is $A$ (we will soon denote by $R$
the area of a rectangular city, and by $C$ the area of a
circular city), we further define

\begin{dfn}\label{defFour}
The \emph{Infeasibility Factor} ($IF(p)$), corresponding to a particular
departure point $p$ within the area of the city, is the 
area of $I(p)$ relative to the overall city area, expressed
as a percentage, i.e.
\begin{equation}
    IF(p) = (100)
    \frac{\iint_{I(p)}  \,dx\,dy}{\iint_{A}  \,dx\,dy}
\end{equation}
\end{dfn}

\begin{figure}[H]
\centering
\begin{subfigure}[b]{0.490\textwidth}
    \centering
    \includegraphics[width=\textwidth]{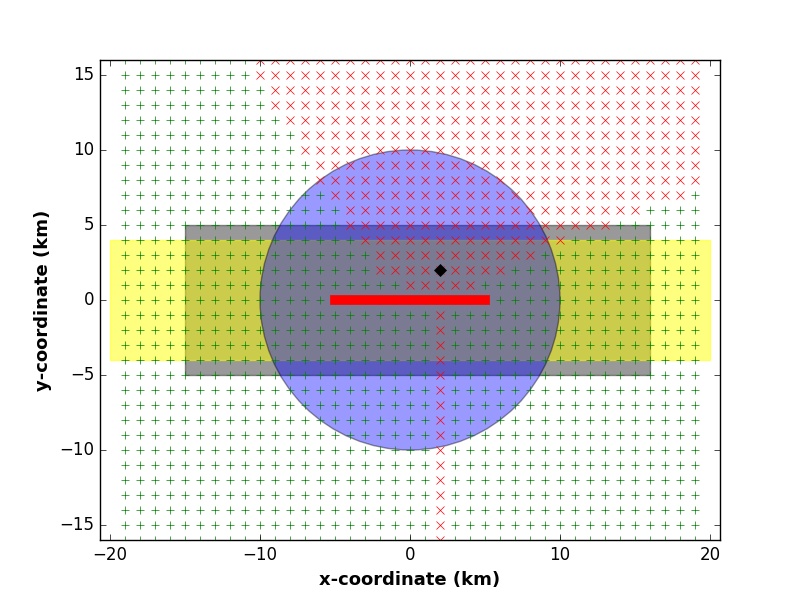}
    \caption[Network2]%
    {{\small Departure point $p = (2,2)$: $IF(p)
    \approx 9/15/35 \%$ for the $l=4, l=3$ and circle cases
    respectively (see Definition \ref{defFour}).}}    
\end{subfigure}
\hfill
\begin{subfigure}[b]{0.490\textwidth}  
    \centering 
    \includegraphics[width=\textwidth]{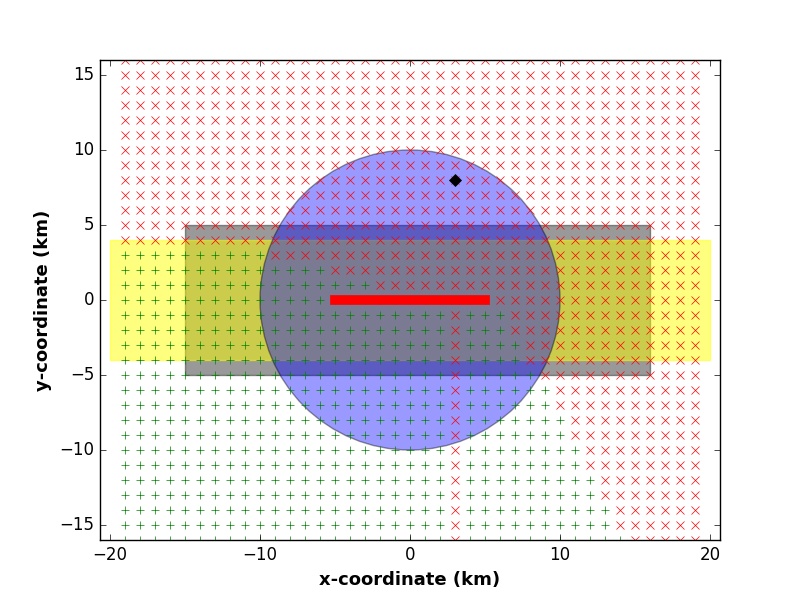}
    \caption[]%
    {{\small Departure point $ p = (3,8)$: $IF(p)
    \approx 51 \%$ for the circle.}}    
\end{subfigure}
\vskip\baselineskip
\begin{subfigure}[b]{0.475\textwidth}   
    \centering 
    \includegraphics[width=\textwidth]{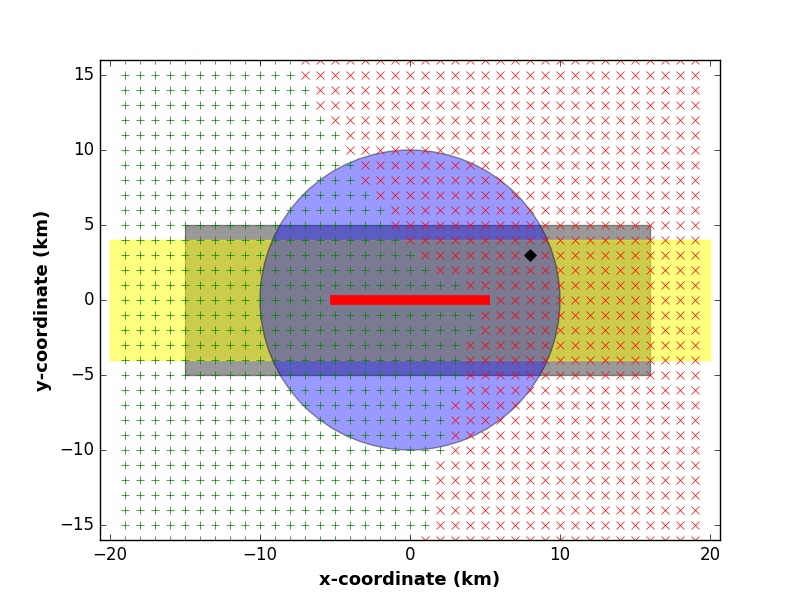}
    \caption[]%
    {{\small Departure point $ p =  (8,3)$: $IF(p)
    \approx 40/38/42 \%$ for the $l=4, l=3$ and circle cases respectively.}}    
\end{subfigure}
\quad
\begin{subfigure}[b]{0.475\textwidth}   
    \centering 
    \includegraphics[width=\textwidth]{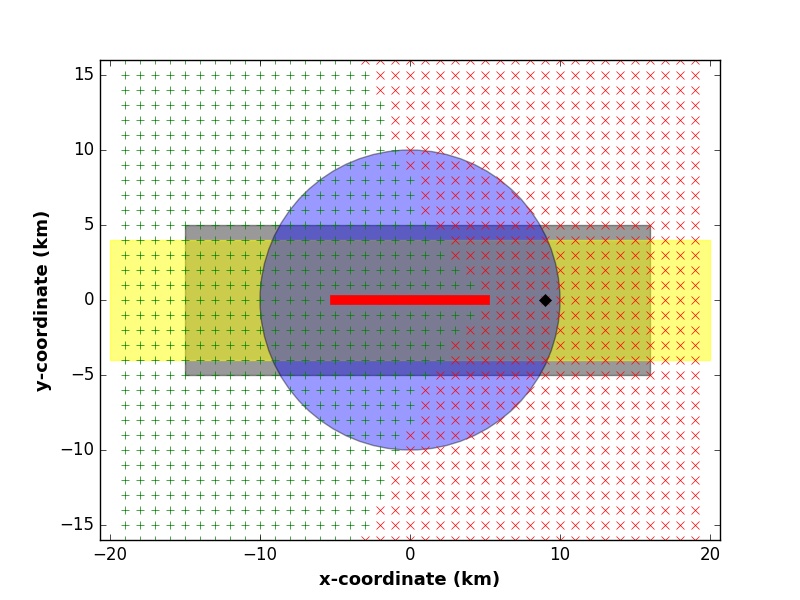}
    \caption[]%
    {{\small Departure point $p = (9,0)$: $IF(p)
    \approx 40/37/39 \%$ for the $l=4, l=3$ and circle cases respectively.}}    
\end{subfigure}
\caption{\small
Infeasible regions for four different (departure) points. In each image, the point is a bold black dot and the tram line is
a solid (red) line centered at $(0,0)$. Three different city shapes, of equal area (about 300 square kilometres), are indicated: (i) circular (blue), (ii) rectangular with $l=3$ (grey) and (iii) rectangular with 
$l=4$ (yellow). The infeasible region is shaded in (red)
{\tt x} signs. Because of symmetries, our examples are all of points in the upper right quadrant.\label{infeas1}}
\end{figure}

\begin{dfn}
The \emph{Infeasibility Factor} (IF) for the city as a 
whole is the average over all points of the Infeasibility
Factors for each point,
\begin{equation}\label{infeas7}
    IF = 
    \frac{\iint_{A} IF((x,y)) \,dx\,dy}{\iint_{A}  \,dx\,dy},
\end{equation}
where we write $p$ as $(x,y)$.
\end{dfn}
In Figure \ref{infeas1} we show examples of infeasible regions for a number points, superimposed on a circular city and two different rectangular cities. In Appendix \ref{appOne} we present further plots showing infeasible regions for various values of $\alpha$ and $\beta$.

.
\subsection{The shape of Infeasible Regions}\label{sec:Shape}
The boundary between the infeasible and feasible regions is defined by the set of points $p$ 
satisfying $d_{\rm euc}(p,p_1) = d_{\rm tr}(p,p_1)$, where $p_1$ is the departure point.
We suppose again our departure point $p_1 = (x_1,y_1)$ is in the
positive quadrant (i.e.\ $x_1>0$ and $y_1>0$). We distinguish
 two cases:
 \subsubsection{Departure points ``along'' the tram line.}
    Here, $0\leq x_1\leq a$.
We set $p_1=(x_1,y_1), p = (x,y)$. Since $0\leq x_1\leq a$, $x_1^t = x_1$. Since $\alpha = \infty$, the second
    term in Equation (\ref{trMetric}) is zero, so we have
    \begin{equation}
        d_{\rm tr} (p_1,p) = 
    d_{\rm euc} ((x_1,y_1), (x_1,0))
    + d_{\rm euc} (Proj(p), p) = y_1 + d_{\rm euc} (Proj(p), p).
    \end{equation}
    This leads us to three further cases:
    \begin{description}

    \item[\bf $\bf x < -a$:]
    In this region, $Proj(p) = (-a,0)$ (i.e.\ $(-a,0)$ is the point on the tram line closest to $(x,y)$), so 
    we have 
    \begin{equation}
        d_{\rm tr} (p_1,p) = 
  y_1 + d_{\rm euc} ((-a,0), (x,y)) = y_1 + \sqrt{(x+a)^2 + y^2}.
    \end{equation}
    The infeasible region is thus bounded by the curve
        \begin{equation}
        d_{\rm euc}(p,p_1) = d_{\rm tr}(p,p_1) \implies 
        \sqrt{(x-x_1)^2 + (y-y_1)^2} 
        = y_1 + \sqrt{(x+a)^2 + y^2}
    \end{equation}
    which can be re-written as 
    \begin{equation}
    4[(x_1+a)^2-y_1^2]x^2
    +8y_1(x_1+a)xy
    +4[(a+x_1)^2(a-x_1)-2ay_1^2]x
    +4y_1(a^2-x_1^2)y
    +(x_1^2-a^2)^2-4y_1^2a^2
    =0.
    \end{equation}
    This equation is of the form $Ax^2 + Bxy + Cy^2 + Dx + Ey + F = 0$ with $C=0$. Since 
    $B^2-4AC = B^2 \geq 0$, it is hyperbolic unless $y_1 = 0$, in which case it is parabolic.

    \item[$\bf -a \leq x < a$:]
    $Proj(p) = (x,0),\ d_{\rm tr} (p_1,p) = 
    y_1 + d_{\rm euc} (Proj(p), p) = y_1 + |y|$ which means we have the curve
    \begin{equation}\label{eqPara}
    d_{\rm euc}(p,p_1) = d_{\rm tr}(p,p_1) \implies 
            \sqrt{(x-x_1)^2 + (y-y_1)^2} 
        = y_1 + |y|.
    \end{equation}
    For negative $y$, this is the straight line $x=x_1$ (visible in Figure \ref{infeas1} (a) and (b)),
    while for positive $y$ it gives
    $y = (x-x_1)^2/(4y_1)$, which is a parabola with focus $(x_1,y_1)$
    and directrix $y=-y_1$.

    \item[$\bf a \leq x $:]
    $Proj(p) = (a,0)$ (i.e.\ $(a,0)$ is the point on the tram line closest to $(x,y)$), so 
    we have 
    \begin{equation}
        d_{\rm tr} (p_1,p) = 
  y_1 + d_{\rm euc} ((a,0), (x,y)) = y_1 + \sqrt{(a-x)^2 + y^2}.
    \end{equation}
        The infeasible region is thus bounded by the curve
        \begin{equation}
        d_{\rm euc}(p,p_1) = d_{\rm tr}(p,p_1) \implies 
        \sqrt{(x-x_1)^2 + (y-y_1)^2} 
        = y_1 + \sqrt{(a-x)^2 + y^2}
    \end{equation}
        which can be re-written as 
    \begin{equation}
    4[(a-x_1)^2-y_1^2]x^2
    -8y_1(a-x_1)xy
    -4[(a-x_1)^2(a+x_1)-2ay_1^2]x
    +4y_1(a^2-x_1^2)y
    +(a^2-x_1^2)^2-4y_1^2a^2
    =0.
    \end{equation}
    This is hyperbolic unless either $y_1=0$ or $x_1=a$, in which case it is parabolic.
    \end{description}
These three curves intersect at the two points $(-a,(a+x_1)^2/(4y_1))$ and $(a,(a-x_1)^2/(4y_1))$.
The asymptotic behaviour ``to the left'' (as $x\to -\infty$) and ``to the right''
(as $x\to \infty$) is as follows:
\begin{description}
\item[\bf As $\bf x\to -\infty$: ]
The 
hyperbola to the left has asymptotic slope $(\phi^+ - 1/\phi^+)/2$, where 
$\phi^+ = y_1/(x_1+a)$. This means
\begin{equation}
     \mbox{the asymptotic slope as } x\to -\infty \mbox{ is }
     \begin{cases}
        \text{positive}  & \text{when } y_1 > x_1 + a\\
        \text{zero}  & \text{when } y_1 = x_1 + a \\
        \text{negative}  & \text{when } y_1 < x_1 + a.
    \end{cases}
\end{equation}
\item[\bf As $\bf x \to \infty$: ]
 The hyperbola to the right has asymptotic slope 
$(\phi^- - 1/\phi^-)/2$, where 
$\phi^- = y_1/(x_1-a)$. This means
\begin{equation}
     \mbox{the asymptotic slope as } x\to \infty \mbox{ is }
     \begin{cases}
        \mbox{positive}  & \mbox{when } y_1 < a - x_1 \\
        \mbox{zero}  & \mbox{when } y_1 = a - x_1  \\
        \mbox{negative}  & \mbox{when } y_1 > a - x_1.
    \end{cases}
\end{equation}
\end{description}
\begin{figure}[ht!]
\centering
\includegraphics[width=12cm, trim={0 0.5cm 0 0},clip]{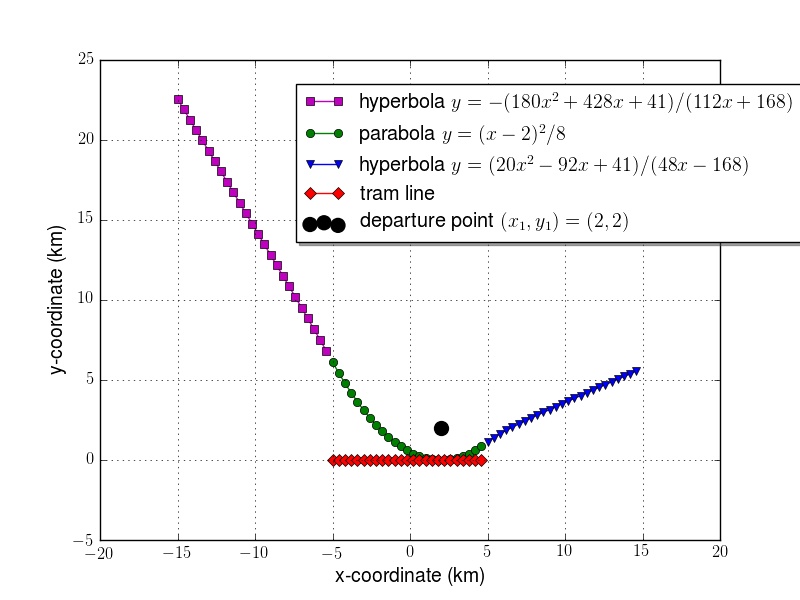}
\caption{\small Boundary of the Infeasible 
Region for departure point $(x_1,y_1) = (2,2)$
(see also Figure \ref{infeas1}(a)). 
The parabola joins with the
hyperbola to the left at $(-5,6.125)$ and with
the hyperbola
to the right at $(5,1.125)$.
Asymptotically, the hyperbolas to the left/right are straight lines with 
slopes $-45/28$ and $5/12$ respectively.
\label{fig:bound1}}
\end{figure}
\subsubsection{Departure points ``beyond'' the tram line}
Here, $x_1 > a$.
We have that $x_1^t = a$, so $Proj(p_1) = (a,0)$. Equation (\ref{trMetric}) gives us
    \begin{equation}
        d_{\rm tr} (p_1,p) = 
    d_{\rm euc} ((x_1,y_1), (a,0))
    + d_{\rm euc} (Proj(p), p) = \sqrt{(x_1-a)^2 + y_1^2} + d_{\rm euc} (Proj(p), p).
    \end{equation}
This leads us to three further cases:
\begin{description}
\item[\bf $\bf x < -a$:] 
$Proj(p) = (-a,0)\ \implies$
    \begin{equation}
        d_{\rm tr} (p_1,p) = 
\sqrt{(x_1-a)^2 + y_1^2}  + d_{\rm euc} ((a,0), (x,y)) = \sqrt{(x_1-a)^2 + y_1^2}   + \sqrt{(x+a)^2 + y^2}.
    \end{equation}
The infeasible region is bounded by the curve
        \begin{equation}
        d_{\rm euc}(p,p_1) = d_{\rm tr}(p,p_1) \implies 
        \sqrt{(x-x_1)^2 + (y-y_1)^2} 
        = \sqrt{(x_1-a)^2 + y_1^2}   + \sqrt{(x+a)^2 + y^2}.
    \end{equation}
After repeated squaring, this can be re-written as 
$Ax^2 + Bxy + Cy^2 + Dx + Ey + F = 0$
where 
\begin{align}
A &= 4ax_1-y_1^2
&
B &= 2y_1(x_1+a)
&
C &= -(x_1-a)^2\nonumber\\
D &= 2a(2ax_1-2x_1^2-y_1^2)
&
E &= -2ay_1(x_1-a)
&
F &= -a^2y_1^2
\end{align}
Since $B^2-4AC = 16ax_1(y_1^2 + (x_1-a)^2) > 0$ (except for the particular point $p_1=(a,0)$), this is a hyperbola.

\item[\bf $\bf -a \leq x < a$:]
$Proj(p) = (x,0)\ \implies$
    \begin{equation}
        d_{\rm tr} (p_1,p) = 
\sqrt{(x_1-a)^2 + y_1^2}  + d_{\rm euc} ((x,0), (x,y)) = |y| + \sqrt{(x_1-a)^2 + y_1^2}.
    \end{equation}
The infeasible region is bounded by the curve
\begin{equation}
d_{\rm euc}(p,p_1) = d_{\rm tr}(p,p_1) \implies 
\sqrt{(x-x_1)^2 + (y-y_1)^2} 
= |y| + \sqrt{(x_1-a)^2 + y_1^2}.
\end{equation}
This gives us
\begin{equation}
x^2  -2x_1x -2\left(y_1\pm \sqrt{(x_1-a)^2 + y_1^2}\right)y  +a(2x_1-a) = 0.
\end{equation}
Since $B=C=0$, this is a parabola.
\item[\bf $\bf a \leq x$:]
$Proj(p) = (a,0)$ so the equation 
$d_{\rm euc}(p,p_1) = d_{\rm tr}(p,p_1)$ reads
\begin{equation}
d_{\rm euc}(p,p_1) = d_{\rm euc}(p,(a,0)) + d_{\rm euc}((a,0), p_1).
\end{equation}
Because of the triangle inequality, this equation has no solution except for the
single point $p=(a,0)$.
\end{description}

\subsection{Infeasibility Factor for a square city}
As before, we let our line run between points $(-a,0)$ and 
$(a,0)$ through the center of a square city.
The square city (of area $4a^2$) is bounded by the four vertices 
$(a,a), (a,-a), (-a,a)$ and $(-a,-a)$. (In terms of our original parameters in 
Section \ref{sec:Rect}, $l=1$ and $p=a$.) We prove that this setup has 
an Infeasibility Factor of $$\mbox{IF} = \frac{7}{72} + \frac{\ln{2}}{3} \approx 0.328$$ (just under one third).
For the technical details of this exact calculation, see Appendix \ref{appexactIF}.

\subsection{Asymptotic Infeasibility for a rectangular city}
For comparison with the square city case, it is easy to show the IF for a rectangular city that
grows arbitrarily long (and narrow), with constant area, is 0.5. The length of the city is
$lp$, its area is $A = 4a^2$, and the tram is of length $2a$ centered at the center of the city.
We divide the city in to three regions:
\clearpage
\begin{description}
\item[R1] Points $(x,y)$ with $x < -a$
\item[R2] Points $(x,y)$ with $-a \leq x \leq a$
\item[R3] Points $(x,y)$ with $x > a$
\end{description}

We consider the following lower and upper bounds on IF:
\begin{description}
\item[Lower Bound]
A lower bound on the infeasible trips is the set of trips from R1 to R1 and from R3 to R3 (which 
are clearly infeasible). R1 and 
R3 have equal area $(A-2ap)/2$ while R2 has area $2ap$. This means the proportion of trips from
R1 to R1 is $(A-2ap)^2/4A^2$, and likewise for trips from R3 to R3. So, our lower bound is 
$(1 - 2ap/A)^2/2$
\item[Upper Bound]
Since trips from R1 to R3, or from R3 to R1 are clearly feasible, an upper bound is the set of all 
trips excluding these. This gives us an upper bound of 
$2ap/A + 2[(A-2ap)/2A][(A+2ap)/2A] = 1/2 +2ap/A - 2a^2p^2/A^2$
\end{description}
We take the limit as $p\to 0$ of both bounds ($a$ and $A$ are constants) to get an IF of 0.5.
\subsection{Infeasibility in circular and rectangular cases}

We carried out the calculations for Figure \ref{infeas1} by 
discretizing a grid of $ 40 \times 40 $ nodes, and writing
code in PYTHON, using equation (\ref{infeas2}), to determine
infeasible journeys.

\begin{figure}[H]
\centering
\includegraphics[width=18cm, trim={0 2.5cm 0 4cm},clip]{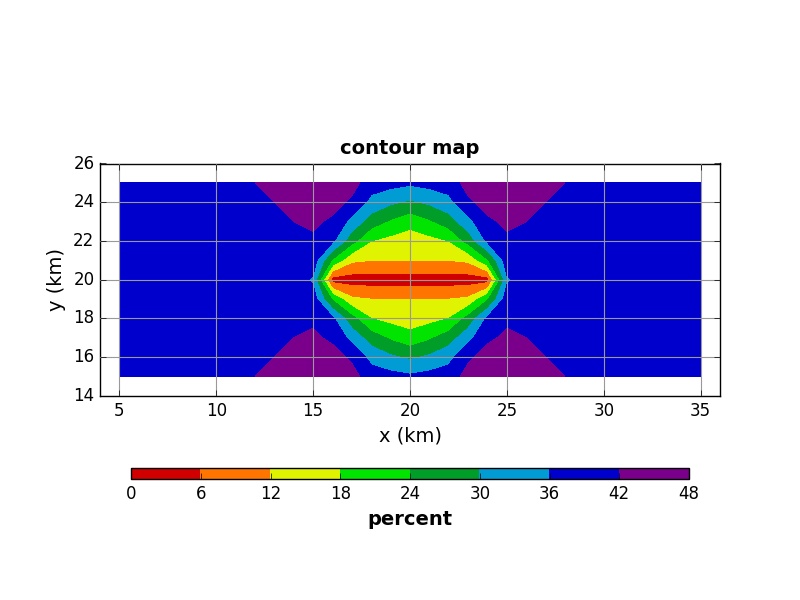}
\caption{\small Infeasibility Factors for 
a rectangular city with $a=5, l=3$.
\label{fig:rFeas}}
\end{figure}

\begin{figure}[H]
\centering
\includegraphics[width=17cm, trim={0 1cm 0 0},clip]{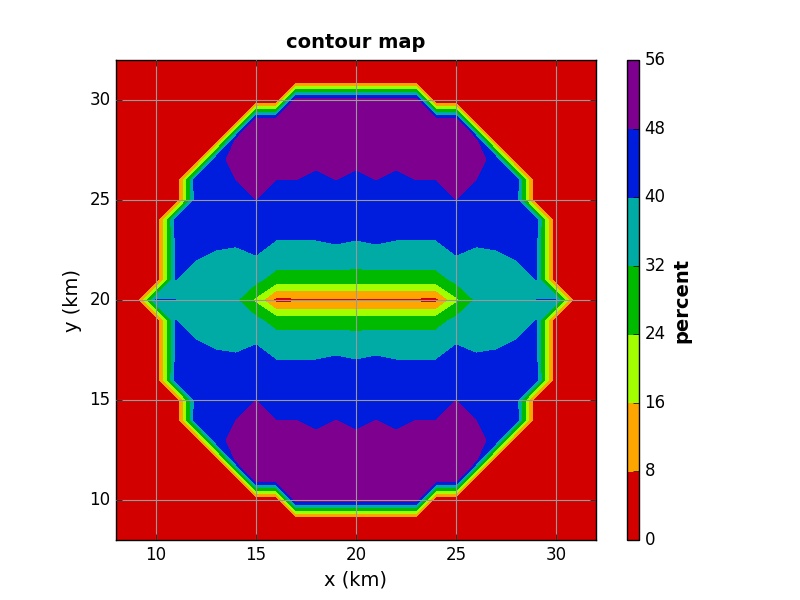}
\caption{\small Infeasibility Factors for 
a circular city with $a=5$.
\label{fig:cFeas}}
\end{figure}

For each city shape, we calculate the Infeasibility Factor for each departure point $p = (x,y)$. 
From Figure \ref{infeas1}, this
corresponds to the area marked by (red) {\tt x} signs 
within the city, divided by the total city area.
In Figure \ref{fig:rFeas} we show the dependence of 
$IF(p)$ on the position $p = (x,y)$ for a rectangular
city with $l=3$. Figure \ref{fig:cFeas} presents the corresponding results for a circular city.

.

The Infeasibilty Factors, for a range of realistic values of $\alpha$ and $\beta$, in the circular and 
rectangular cases, are presented 
in Appendix \ref{appTwo}. From Figures \ref{fig:infeas4} and \ref{fig:infeas5} we note
\begin{theorem}
The infeasibility factors for the circular case
are higher in all cases (except for Figures
\ref{fig:infeas4} (f) and 
\ref{fig:infeas5} (f), where they are identical)
than for the rectangular case. The difference is particularly
noticeable comparing Figure
\ref{fig:infeas4} (b), (c), (d), (e)
with
\ref{fig:infeas5} (b), (c), (d), (e),
where it is over $20\%$.
\end{theorem}

\begin{figure}[ht!]
\centering
\begin{subfigure}[b]{0.410\textwidth}
    \centering
    \includegraphics[width=\textwidth, trim={0 2.5cm 7cm 0},clip]{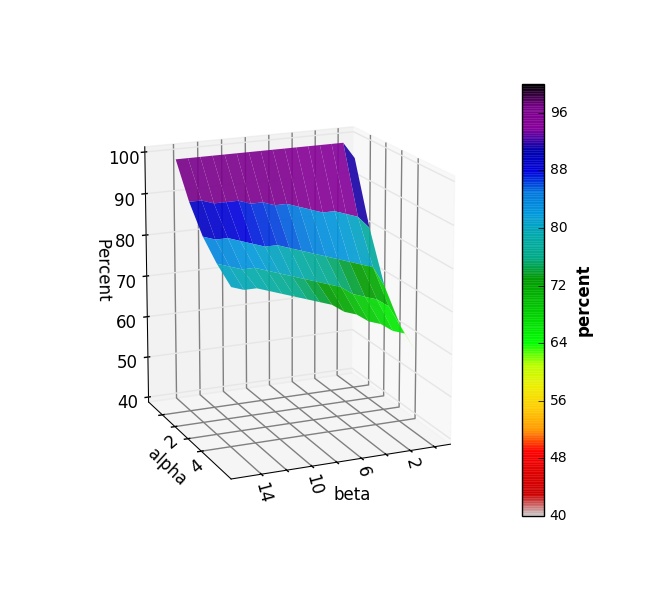}
    \caption[Network2]%
    {{\small Circular Case}}
\end{subfigure}
\hfill
\begin{subfigure}[b]{0.570\textwidth}  
    \centering 
    \includegraphics[width=\textwidth, trim={0 2.5cm 0 0},clip]{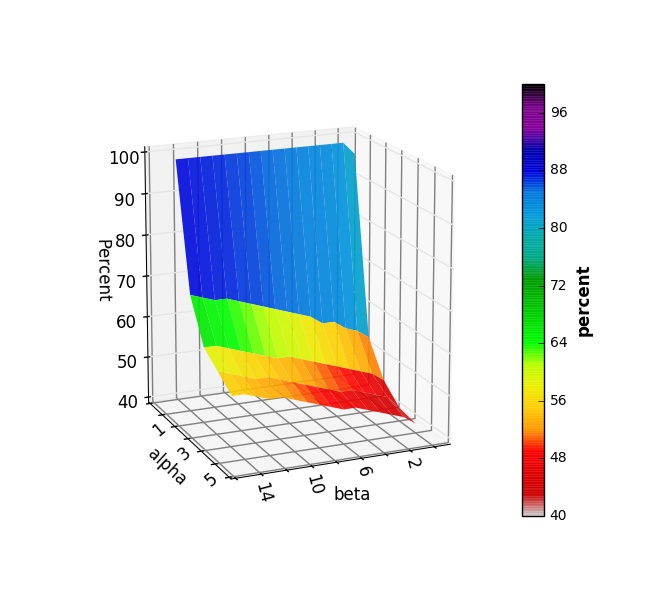}
    \caption[]%
    {{\small Rectangular Case}}    
\end{subfigure}
\caption{\small
Overall Infeasibility Factors for the circular case (a) and rectangular case (b) as they depend on the tram speed (fixed by $\alpha$) and frequency of service (fixed by $\beta$).
\label{final1}}
\end{figure}

We discretize equation \ref{infeas2} to calculate a single number (percentage) for each city shape, representing an 
average over $p$ of all $IF(p)$. This number still depends on
\begin{itemize}
    \item $\alpha$
    \item $\beta$
    \item $a$ (fixed by the length of the tram line)
    \item the city shape
\end{itemize}
We fix $a=r/2$ and present in Figure \ref{final1} results for the Infeasibility Factor
(IF) as it depends on 
$\alpha$ and $\beta$ for the rectangular and circular cases. 
.
\begin{theorem}
Note from Figure \ref{final1} the lower overall Infeasibility Factors for the
rectangular case: much of the graph is around 
$55\%$ (yellow), while for the circular case, much 
of the graph is around $85\%$ (blue).
\end{theorem}
This 
plot should allow a designer / transport engineer, who will have 
an estimate for values of $\alpha$ and $\beta$, to see which 
IF corresponds to their system, given the city geometry.
\section{Case Study: Galway City}
We present in Appendix \ref{appGal} a case study for Galway City (Ireland), a rectangular city whose 
``length'' is about 3 times its ``width'' ($l=3$).
\section{Conclusions}
We have presented a model here that enables mathematical calculation of the utility of a light rail / tram line in a city of a certain shape. We have presented in detail the calculations for a single line tram in rectangular and circular cities, showing the lower infeasibility factors for such a tram in the rectangular case: We therefore argue that building such a tram in a rectangular city is more feasible than in a circular city. This is illustrated in the case study for Galway City.

For further work we envisage the following:
\begin{itemize}
\item
We will investigate 
the dependence of the infeasibility factors, not just on the tram speeds and tram frequencies, but also on the length of the tram line (relative to the city size).
\item
We will construct
a more elaborate model, using the metrics presented here, for more elaborate tram networks with multiple (intersecting) lines. This should model larger real-world city transit systems.
\item
In our calculations, we assume all journeys are equally probable. This assumption could be relaxed, building a model which calculates infeasibility
factors that are weighted averages of different journeys. In real-world scenarios, the probability of going from point $i$ to $j$ depends not just on 
population densities (which in any case will not be uniform), but on other features. For example, one
or other of $i$, $j$ may correspond to a place of work, a hospital, a University, a transport hub, a shopping centre: Journeys to/from these locations may have higher weightings, even though the population densities at these locations may be lower. (Thanks to Ulf Strohmayer for pointing this out.)
\end{itemize}



\clearpage
\begin{appendices}
\input{AInfeasReg}
\input{AexactIF}

\input{AInfeasFact}
\input{AGalway}
\end{center}
\end{appendices}
\end{document}

%% file: AInfeasReg.tex
\section{Examples of ($\alpha$,$\beta$)--infeasible regions}\label{appOne}
\begin{figure}
\centering
\begin{subfigure}[b]{0.490\textwidth}
    \centering
    \includegraphics[width=\textwidth]{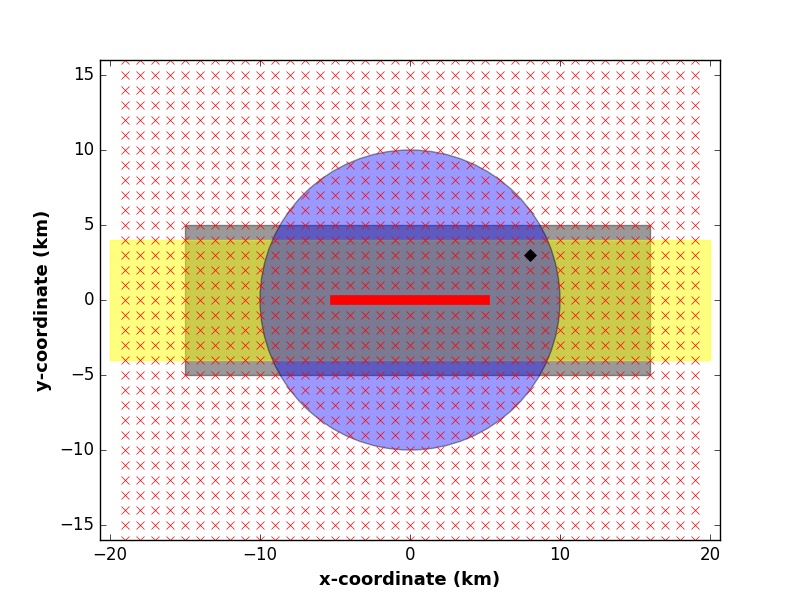}
    \caption[Network2]%
    {{\small $(\alpha,\beta) = 
    (0,\infty)$. $IF((8,3)) = 100\%$ 
    (see Definition \ref{defFour}).}}    
\end{subfigure}
\hfill
\begin{subfigure}[b]{0.490\textwidth}  
    \centering 
    \includegraphics[width=\textwidth]{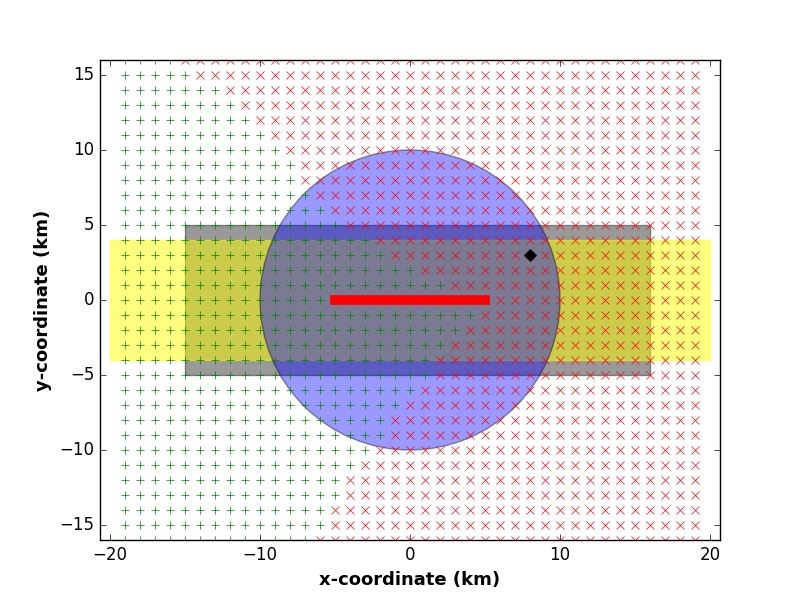}
    \caption[]%
    {{\small $(\alpha,\beta) = 
    (3,10)$. $IF((8,3)) \approx 66/49/48\%$ for the
    circle/$l=3$/$l=4$ cases respectively.}}
\end{subfigure}
\vskip\baselineskip
\begin{subfigure}[b]{0.475\textwidth}   
    \centering 
    \includegraphics[width=\textwidth]{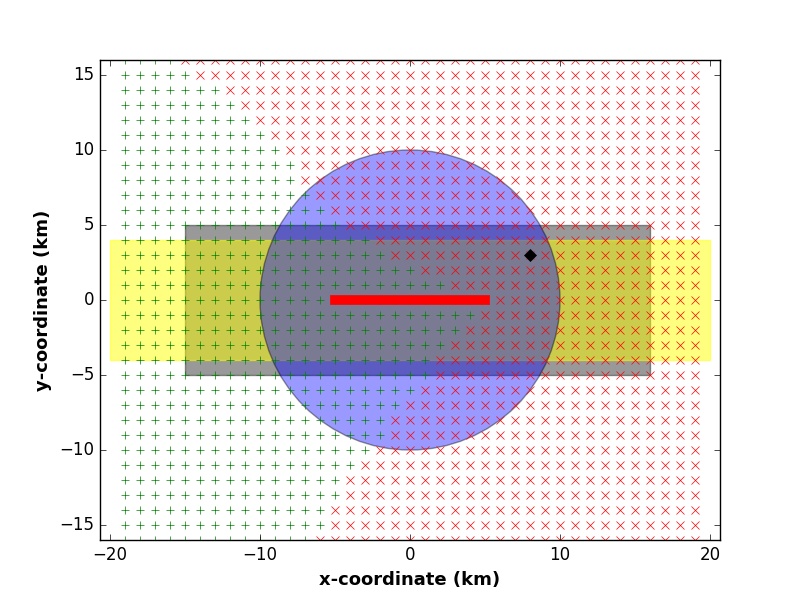}
    \caption[]%
    {{\small $(\alpha,\beta) = 
    (3,5)$. $IF((8,3)) \approx 61/46/45\%$ for the
    circle/$l=3$/$l=4$ cases respectively.}}
\end{subfigure}
\quad
\begin{subfigure}[b]{0.475\textwidth}   
    \centering 
    \includegraphics[width=\textwidth]{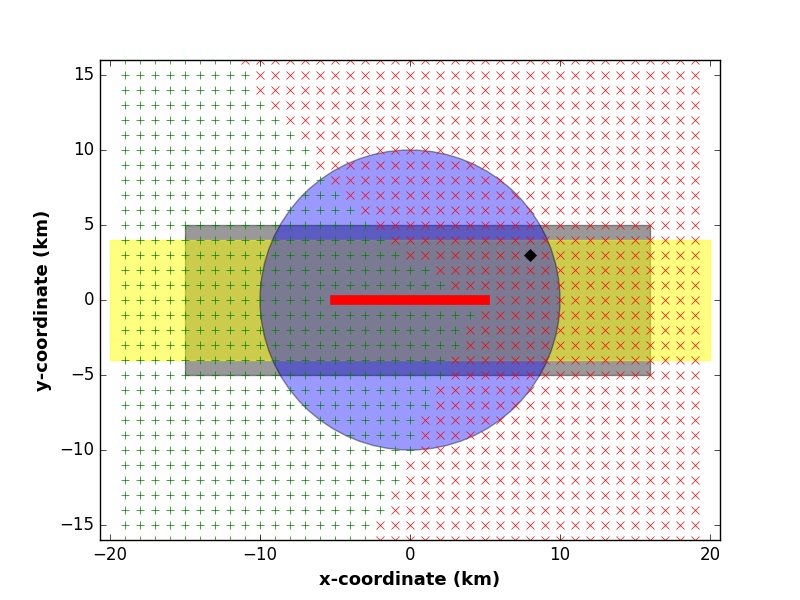}
    \caption[]%
    {{\small $(\alpha,\beta) = 
    (5,10)$. $IF((8,3)) \approx 58/46/45\%$ for the
    circle/$l=3$/$l=4$ cases respectively.}}
\end{subfigure}
\vskip\baselineskip
\begin{subfigure}[b]{0.475\textwidth}   
    \centering 
    \includegraphics[width=\textwidth]{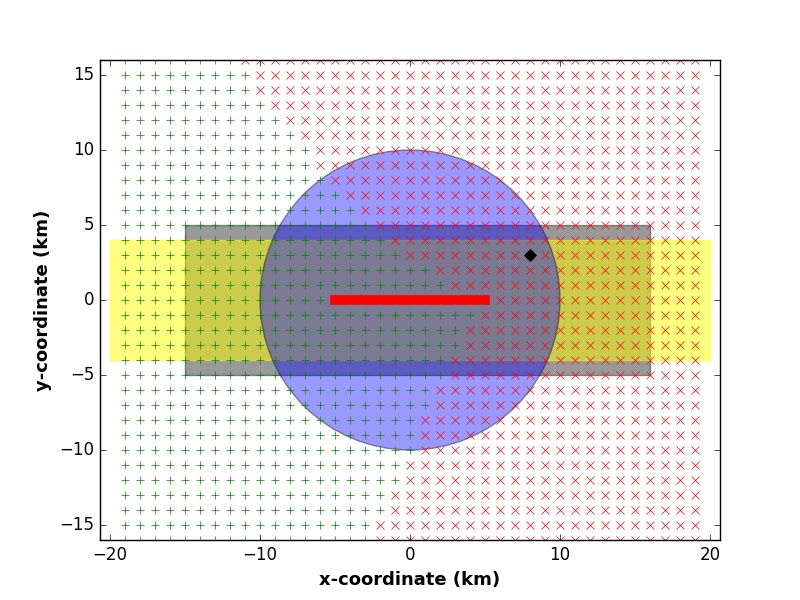}
    \caption[]%
    {{\small $(\alpha,\beta) = 
    (5,5)$. $IF((8,3)) \approx 54/43/43\%$ for the
    circle/$l=3$/$l=4$ cases respectively.}}
\end{subfigure}
\quad
\begin{subfigure}[b]{0.475\textwidth}   
    \centering 
    \includegraphics[width=\textwidth]{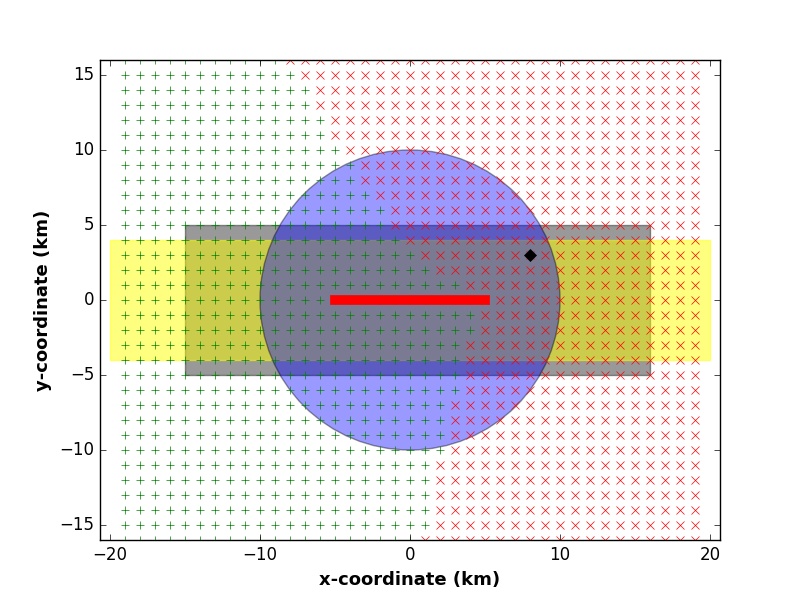}
    \caption[]%
    {{\small $(\alpha,\beta) = 
    (\infty, 0)$. $IF((8,3)) \approx 42/38/40\%$ for the
    circle/$l=3$/$l=4$ cases respectively.}}
\end{subfigure}
\caption{\small
Infeasible regions for the (departure) point $(8,3)$ with six different pairs of $(\alpha,\beta)$ values. The departure point is marked with a bold black dot and the tram line is
a solid (red) line centered at $(0,0)$. Three different city shapes, of equal area (about 300 square kilometres), are indicated: (i) circular (blue), (ii) rectangular with $l=3$ (grey) and (iii) rectangular with 
$l=4$ (yellow). The infeasible region is shaded in (red)
{\tt x} signs.\label{infeas3}}
\end{figure}

To help the reader appreciate the role of the tram speed, and 
the frequency of service, on the infeasible regions, we show
in Figure \ref{infeas3} some further plots of infeasibile regions for some realistic values of $\alpha$ and $\beta$ and departure point $(8,3)$. We remind the reader 
that
$\alpha = \Bar{s}_{\rm t} / \Bar{s}_{\rm nt}$ measures the
(relative) tram speed while
$\beta = t_f/2$ measures the frequency. Note
\begin{itemize}
    \item Figure \ref{infeas3} (a) corresponds to the extreme
    case scenario where the tram does not move (its speed is
    zero) and/or the time interval between trams is infinite.
    In this (ridiculous!) situation, obviously the infeasibility factor is $100\%$ (there is no reason to
    take such a tram!)
    \item Plots (c) and (d) of Figure \ref{infeas3} are very
    similar: (d) has higher speed trams, while (c)  has 
    lower speed trams but more frequent ones.
\end{itemize}
.


%% file: AexactIF.tex
\section{Infeasibility Factor for Square City}\label{appexactIF}
Without loss of generality, we restrict ourselves to points in the upper right quadrant,
i.e.\ $0\leq x_1 \leq a$ and $0\leq y_1 \leq a$. For points in this quadrant, the boundary of the 
infeasible region is parabolic (as described in Section \ref{sec:Shape}). Depending on the 
point $(x_1,y_1)$ chosen, the parabolic boundary may intersect the city boundaries along the 
horizontal line $y=a$ or along the vertical lines $x=-a$ or $x=a$. Points that are close to the 
tram line (i.e. small values of $y_1$) will give rise to ``narrow'' parabolas (i.e. with 
small latus rectum), while larger values of $y_1$ will give ``wider'' parabolas.
There are three possibilities:
\begin{enumerate}
    \item[(i)] Both intersection points are along $y=a$ (``narrowest'' parabolas);
    \item[(ii)] One intersection point is along $y=a$ and one is along $x=a$;
    \item[(iii)] One intersection point is along $x=-a$ and one is along $x=a$ (``widest'' parabolas).
\end{enumerate}
We examine these three cases separately.
\begin{description}
\item[Case (i)] From Equation (\ref{eqPara}), the parabolic boundary is $y = (x - x_1)^2/4y_1$. The right arm of this 
parabola intersects the vertex $(a,a)$ when $a = (a - x_1)^2/4y_1$, i.e. when $y_1 = (a-x_1)^2/4a$. For fixed $x_1$,
values of $y_1$ less than $(a-x_1)^2/4a$ will give (narrow) parabolas that intersect only with $y=a$. This region is thus 
defined by $0\leq x_1\leq a$ and $0\leq y_1\leq (a-x_1)^2/4a$. The area of the infeasible region (bounded by the 
parabola and the line $y=a$) is 
\begin{equation}
A1(x_1, y_1) = \int_{y=0}^a\int_{x=x_1-2\sqrt{y_1y}}^{x_1+2\sqrt{y_1y}} dx dy = \int_0^a 4\sqrt{y_1y} dy = \frac{8\sqrt{a^3y_1}}{3}.
\end{equation}
The average value of this area is thus
\begin{equation}
\overline{A1} = 
\frac{\int_{x_1=0}^a\int_{y_1=0}^{(a-x_1)^2/4a} A1(x_1, y_1) dy_1 dx_1}{\int_{x_1=0}^a\int_{y_1=0}^{(a-x_1)^2/4a} dy_1 dx_1}
= \frac{a^4/18}{a^2/12} = \frac{2a^2}{3}.
\end{equation}
\item[Case (ii)] In this region, the right arm of the parabola intersects the vertical line $x=a$ 
while the left arm intersects the horizontal line
$y=a$. As $y_1$ increases, the left arm eventually intersects the vertex $(-a,a)$ when $a = (-a-x_1)^2/4y_1$, i.e.
$y_1 = (a+x_1)^2/4a$. This region is thus 
defined by $0\leq x_1\leq a$ and $(a-x_1)^2/4a\leq y_1\leq (a+x_1)^2/4a$. The area of the infeasible region (bounded
below by $y = (x - x_1)^2/4y_1$, to the right by $x=a$ and above by $y=a$) is
\begin{equation}
\begin{split}
A2(x_1, y_1) &= \int_{y=0}^{(a-x_1)^2/4y_1}\int_{x=x_1-2\sqrt{y_1y}}^{x_1+2\sqrt{y_1y}} dx dy
+
\int_{y=(a-x_1)^2/4y_1}^{a}\int_{x=x_1-2\sqrt{y_1y}}^a dx dy\\
&=
a\left(a-\frac{(a-x_1)^2}{4y_1}\right)
+
 \int_{y=0}^{(a-x_1)^2/4y_1} (x_1+2\sqrt{y_1y}) dy
-
\int_{y=0}^a (x_1-2\sqrt{y_1y}) dy\\
&=
(a-x_1)\left(a-\frac{(a-x_1)^2}{4y_1}\right)
+2\sqrt{y_1}
\left(\int_{y=0}^{(a-x_1)^2/4y_1} + \int_{y=0}^a \right) (\sqrt{y}) dy\\
&=
a(a-x_1+4\sqrt{ay_1}/3) - (a-x_1)^3/(12y_1).
\end{split}
\end{equation}
We average to get
\begin{equation}
\overline{A2} = 
\frac{\int_{x_1=0}^a\int_{y_1=(a-x_1)^2/4a}^{(a+x_1)^2/4a} A2(x_1, y_1) dy_1 dx_1}{\int_{x_1=0}^a\int_{y_1=(a-x_1)^2/4a}^{(a+x_1)^2/4a} dy_1 dx_1}
=
\frac{a^4(1+6\ln{2})/9}{a^2/2}
=
\frac{2a^2(1 + 6\ln{2})}{9}.
\hfil
\end{equation}
\end{description}
\begin{description}
\item[Case (iii)] 
Here the right arm of the parabola intersects with the vertical line $x=a$ and the left
parabola arm intersects with $x=-a$ (we have a ``wider'' parabola compared to the other cases).
The region is defined by $0\leq x_1\leq a$ and $(a+x_1)^2/4a\leq y_1\leq a$.
 The area of the infeasible region (bounded
below by $y = (x - x_1)^2/4y_1$, to the right by $x=a$, to the left by $x=-a$ 
and above by $y=a$) is
\begin{equation}
A3(x_1, y_1) = \int_{x=-a}^a\int_{y=(x-x_1)^2/4y_1}^{a} dy dx
= 2a^2 - \frac{1}{4y_1}\int_{x=-a}^a (x-x_1)^2 dx
=2a^2 - \frac{a(a^2 + 3x_1^2)}{6y_1}.
\end{equation}
Averaging gives us
\begin{equation}
\overline{A3} = 
\frac{\int_{x_1=0}^a\int_{y_1=(a+x_1)^2/4a}^{a} A3(x_1, y_1) dy_1 dx_1}{\int_{x_1=0}^a\int_{y_1=(a+x_1)^2/4a}^{a} dy_1 dx_1}
=
\frac{a^4(2+6\ln{2})/9}{5a^2/12}
=
\frac{8a^2(1 + 3\ln{2})}{15}.
\hfil
\end{equation}
\end{description}
Taking the weighted average over the three cases gives
\begin{equation}
    \overline{A} =
    \left(\frac{1}{12}\right) \overline{A1} +
    \left(\frac{1}{2}\right) \overline{A2}  +
    \left(\frac{5}{12}\right) \overline{A3}
    =
    \left(\frac{a^2}{12}\right)\left[ \frac{2}{3} + \frac{4}{3}(1+6\ln{2}) + \frac{8}{3}(1+3\ln{2} \right]
    =
    \left(\frac{a^2}{18}\right)(7 + 24\ln{2}).
\end{equation}
Since the city has area $4a^2$, this gives a final (dimensionless) Infeasibility Factor of
$\overline{A}/(4a^2)$, or 
\begin{equation}
\mbox{IF} = \frac{7}{72} + \frac{\ln 2}{3}.
\end{equation}

%% file: AInfeasFact.tex
\section{Infeasibility Factor Examples}\label{appTwo}
\begin{figure}
\centering
\begin{subfigure}[b]{0.32\textwidth}
    \centering
    \includegraphics[height=10cm,trim={12cm 0 7cm 0},clip]{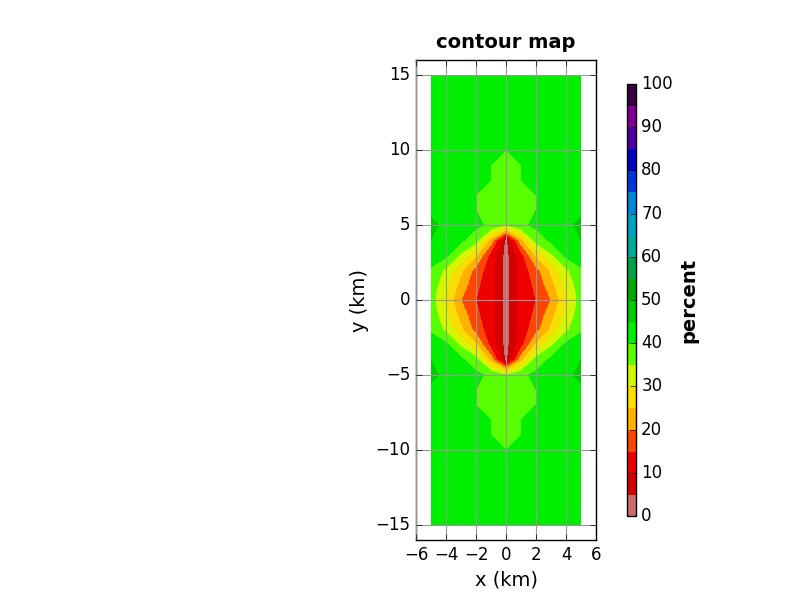}
    \caption[Network2]%
    {{\small \parbox{4cm}{$(\alpha,\beta) = 
    (\infty, 0), IF = 37\%$ }
    }}    
\end{subfigure}
\begin{subfigure}[b]{0.32\textwidth}  
    \centering 
    \includegraphics[height=10cm,trim={12cm 0 7cm 0},clip]{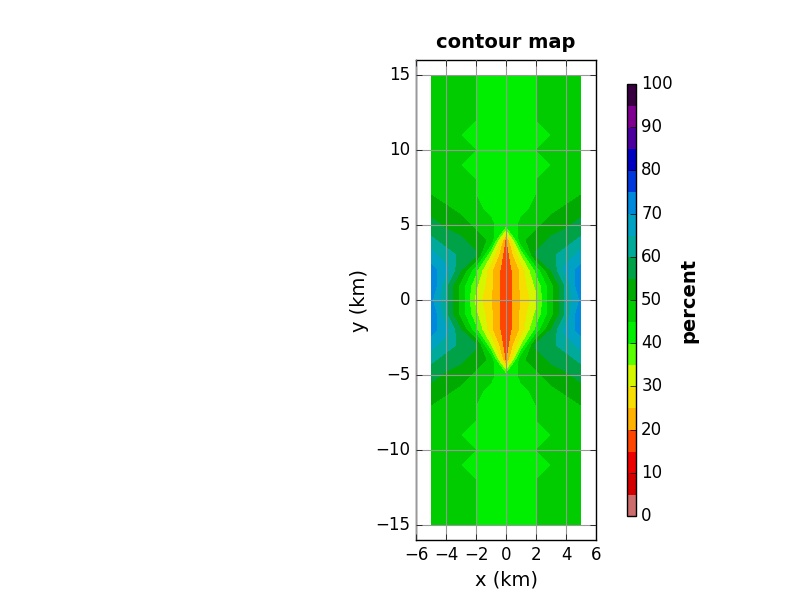}
    \caption[]%
    {{\small \parbox{4cm}{ $(\alpha,\beta) = 
    (5,5), IF = 47\%$}
    }}
\end{subfigure}
\begin{subfigure}[b]{0.32\textwidth}  
    \centering 
    \includegraphics[height=10cm,trim={12cm 0 0 0},clip]{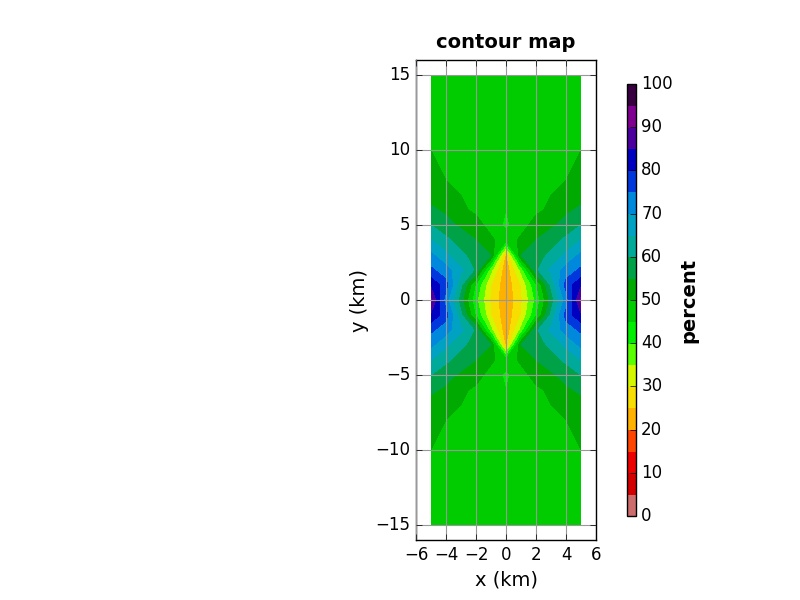}
    \caption[]%
    {{\small \parbox{4cm}{$(\alpha,\beta) = 
    (5,10), IF = 51\%$ }
    }}
\end{subfigure}
\vskip\baselineskip
\begin{subfigure}[b]{0.32\textwidth}  
    \centering 
    \includegraphics[height=10cm,trim={12cm 0 7cm 0},clip]{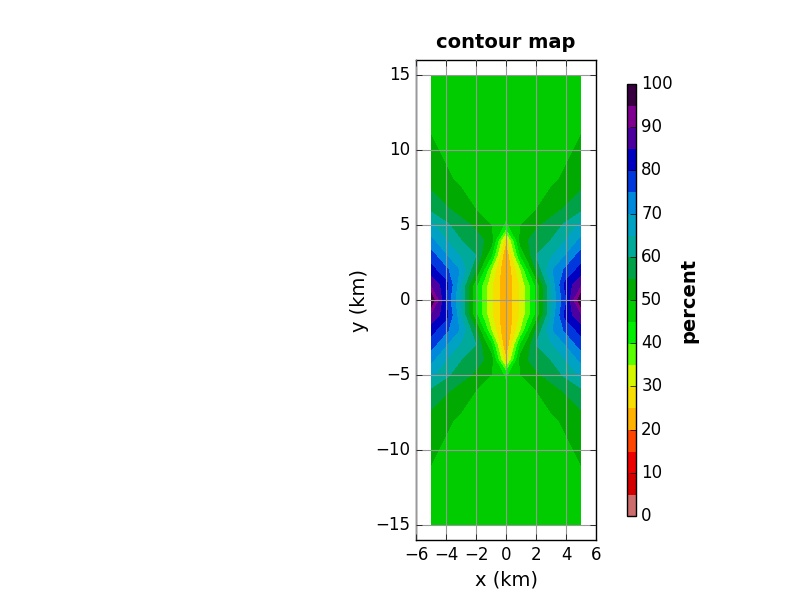}
    \caption[]%
    {{\small \parbox{4cm}{$(\alpha,\beta) = 
    (3,5), IF = 52\%$}
    }}
\end{subfigure}
\begin{subfigure}[b]{0.32\textwidth}  
    \centering 
    \includegraphics[height=10cm,trim={12cm 0 7cm 0},clip]{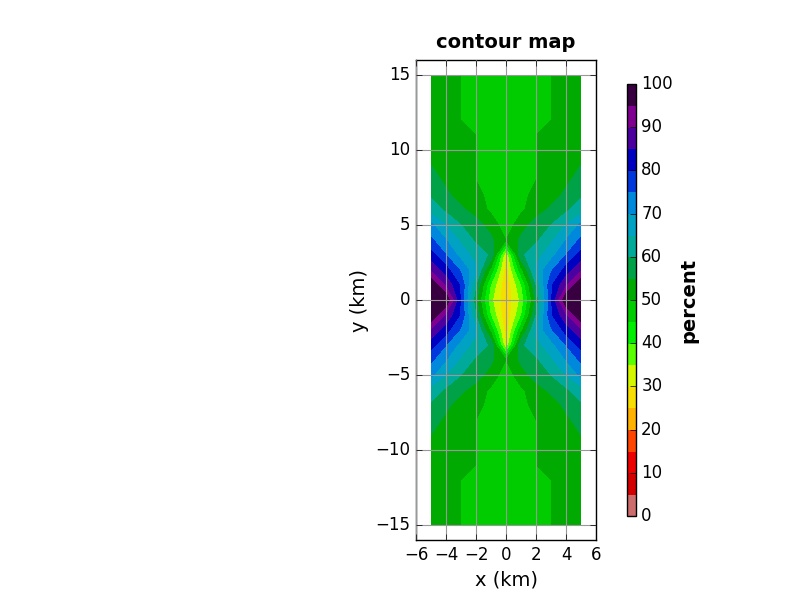}
    \caption[]%
    {{\small \parbox{4cm}{$(\alpha,\beta) = 
    (3,10), IF = 56\%$ }
    }}
\end{subfigure}
\begin{subfigure}[b]{0.32\textwidth}  
    \centering 
    \includegraphics[height=10cm,trim={12cm 0 0 0},clip]{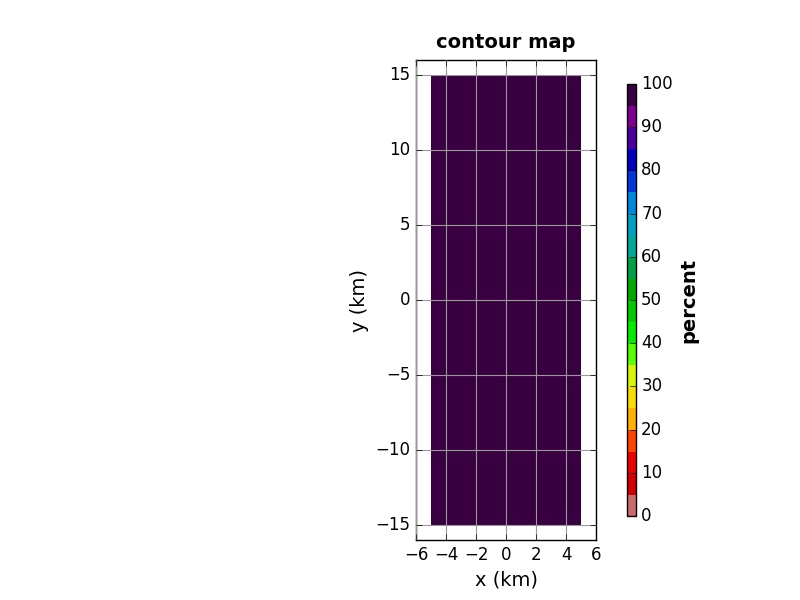}
    \caption[]%
    {{\small \parbox{4cm}{$(\alpha,\beta) = 
    (0,\infty), IF = 100\%$ }
    }}
\end{subfigure}
\caption{\small
Infeasibility Factors (IF) for a rectangular city with
$l=3$ and $a=5$, for some (plausible) values of
$(\alpha, \beta)$. The same color scale (on the right)
is used for all plots.
\label{fig:infeas4}}
\end{figure}

We present in Figure \ref{fig:infeas4} plots of $IF(p)$ for various values of $\alpha$ and $\beta$. Our
city here has area 300 square kilometres, and is a 
rectangle of size $10 \times 30$. Figures 
\ref{fig:infeas4}(a) and \ref{fig:infeas4}(f) present the
extreme case scenarios with fewest/most infeasible 
journeys, respectively.
As we would expect, points at right angles to 
the tram line show highest Infeasibility Factors for
example points $(5,0)$ or $(-5,0)$ in Figures \ref{fig:infeas4} (b),
(c), (d), (e). (Figure \ref{fig:infeas4}(a) is identical 
to Figure \ref{fig:rFeas}, where a different coloring
scheme is used for the contours.)

\begin{figure}
\centering
\begin{subfigure}[b]{0.49\textwidth}
    \centering
    \includegraphics[height=7cm,trim={0 0 7cm 0},clip]{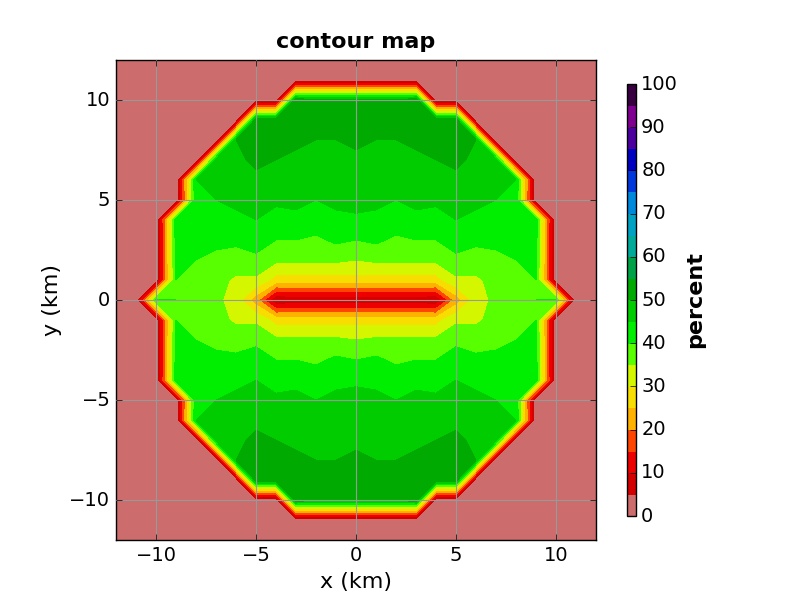}
    \caption[Network2]%
    {{\small \parbox{4cm}{$(\alpha,\beta) = 
    (\infty, 0), IF = 42\%$ }
    }}    
\end{subfigure}
\begin{subfigure}[b]{0.49\textwidth}  
    \centering 
    \includegraphics[height=7cm,trim={0 0 0 0},clip]{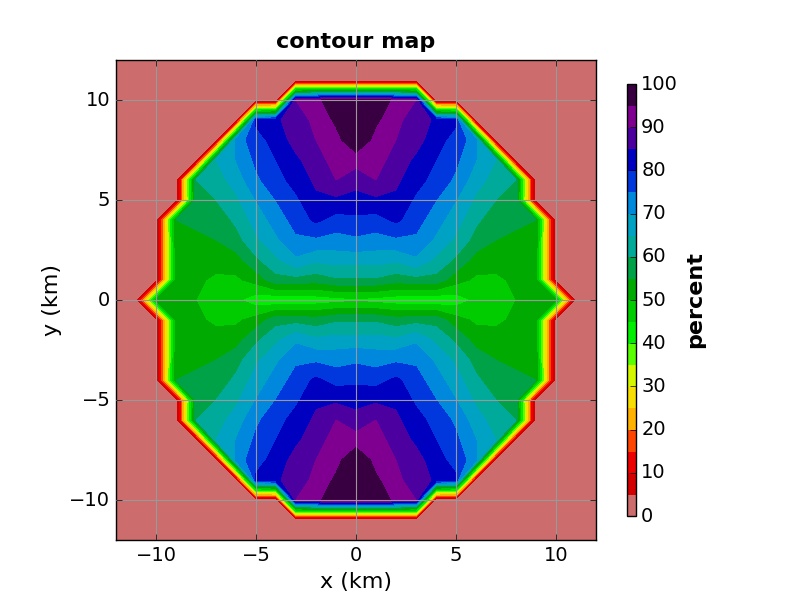}
    \caption[]%
    {{\small \parbox{4cm}{ $(\alpha,\beta) = 
    (5,5), IF = 69\%$}
    }}
\end{subfigure}
\vskip\baselineskip
\begin{subfigure}[b]{0.49\textwidth}  
    \centering 
    \includegraphics[height=7cm,trim={0 0 7cm 0},clip]{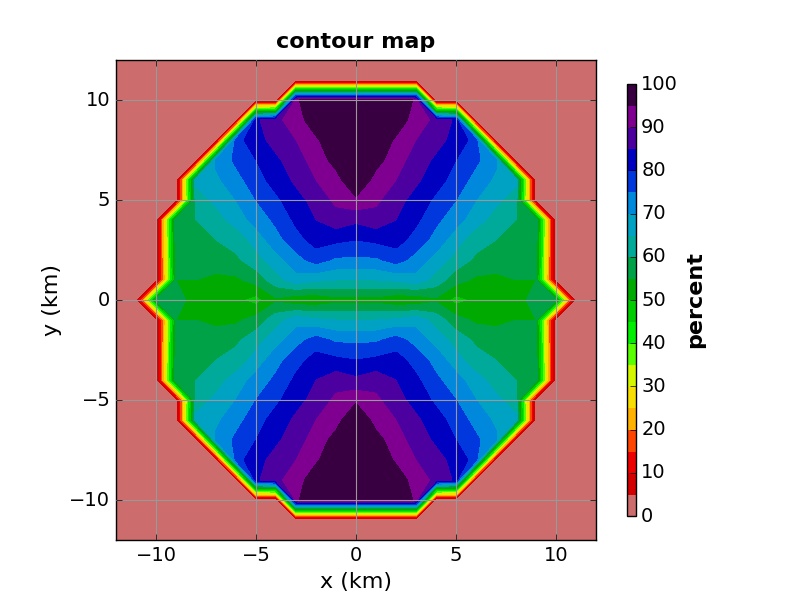}
    \caption[]%
    {{\small \parbox{4cm}{$(\alpha,\beta) = 
    (5,10), IF = 75\%$ }
    }}
\end{subfigure}
\begin{subfigure}[b]{0.49\textwidth}  
    \centering 
    \includegraphics[height=7cm,trim={0 0 7cm 0},clip]{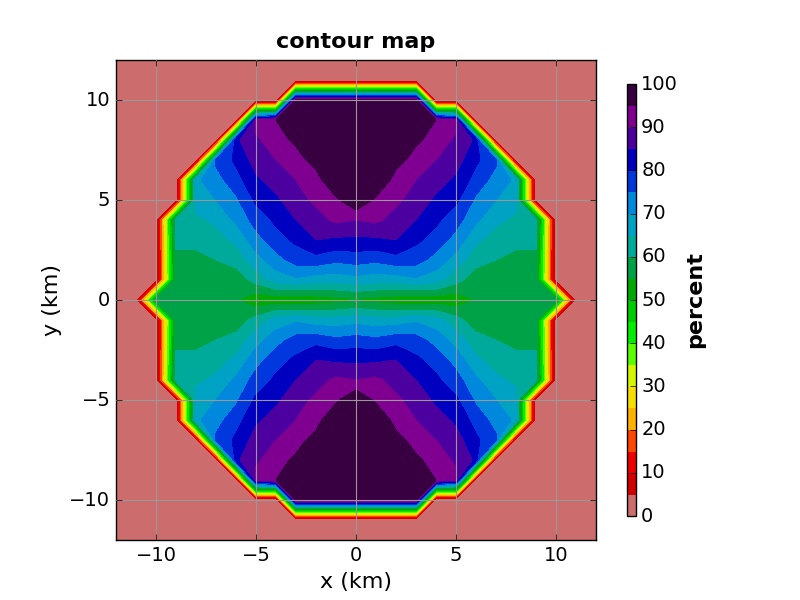}
    \caption[]%
    {{\small \parbox{4cm}{$(\alpha,\beta) = 
    (3,5), IF = 78\%$}
    }}
\end{subfigure}
\vskip\baselineskip
\begin{subfigure}[b]{0.49\textwidth}  
    \centering 
    \includegraphics[height=7cm,trim={0 0 7cm 0},clip]{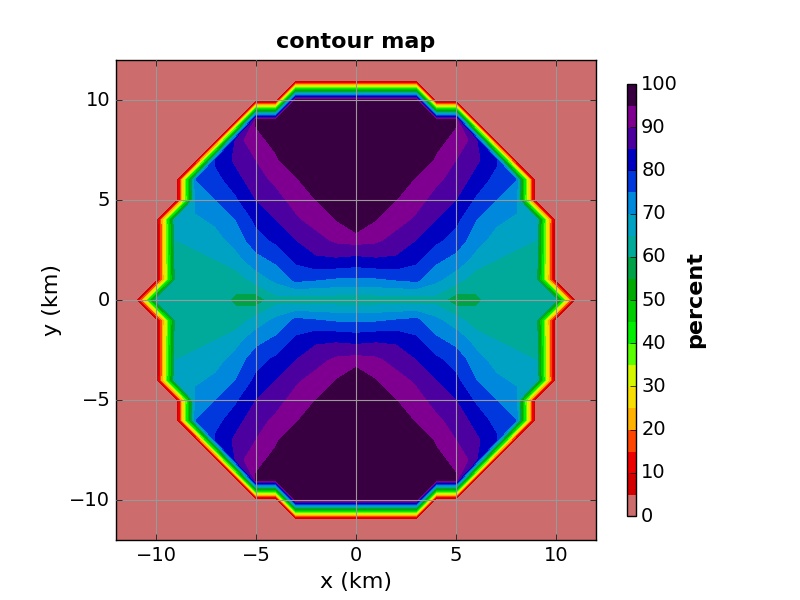}
    \caption[]%
    {{\small \parbox{4cm}{$(\alpha,\beta) = 
    (3,10), IF = 82\%$ }
    }}
\end{subfigure}
\begin{subfigure}[b]{0.49\textwidth}  
    \centering 
    \includegraphics[height=7cm,trim={0 0 7cm 0},clip]{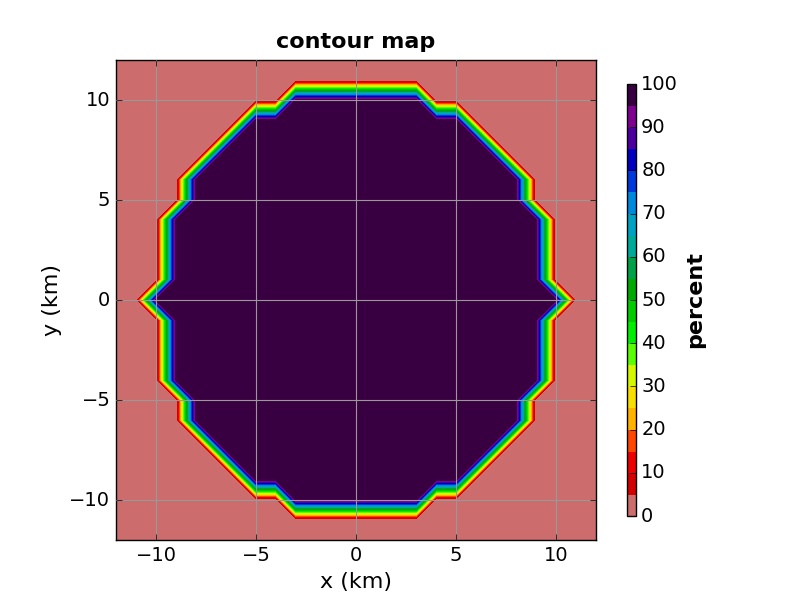}
    \caption[]%
    {{\small \parbox{4cm}{$(\alpha,\beta) = 
    (0,\infty), IF = 100\%$ }
    }}
\end{subfigure}
\caption{\small
Infeasibility Factors (IF) for a circular city with
radius $r=10$. The tram line runs between points 
$(-5,0)$ and $(5,0)$.
 The same color scale (on the right of (b))
is used for all plots.
\label{fig:infeas5}}
\end{figure}

In Figure \ref{fig:infeas5} we present the corresponding 
contour plots for a circular city of approximately equal area (radius 
10km).

.
.

\clearpage

%% file: AGalway.tex
\section{Case study of Galway City}\label{appGal}

We present here a few generic\footnotemark  \ schematic\footnotemark[\value{footnote}] suggestions for the layout of a single line
light rail/tram for Galway City.
\footnotetext{We do not consider any details of city topography, road layout, physical
grography (river, etc.\ ) here, and leave it to others to take these in to account.
Because our layout models link square kilometer areas, they do not have fine grain
detail, and so leave room for the actual line to be placed within hundreds of 
metres of the centre of each square in our grid, without affecting substantially the
details of our calculations.}
Each suggestion differs by taking in to account in
greater detail the variations of population density, and building a 
successively longer line. So in Figure~\ref{app1} our (short) line just links the 
highest population densities (from Figure~\ref{map1}), leading eventually to
Figure~\ref{app5} which is the longest line, taking in to account all the data.

Figure \ref{overGal} shows the overall 
population densities, drawn to scale, 
with some important points of interest
indicated.
Figure \ref{googleG} shows the single
tram line, connecting highest density 
areas, superimposed on a GOOGLE map of
Galway City.
Figure \ref{blueMap} presents data from the CSO (Central Statistics Office,
see \cite{cso1}), via AIRO (All-Island Research Observatory, see 
\cite{airo}), showing the linear/rectangular nature of (the population distribution of) Galway City.

\begin{center}
\input{map1}
\input{data1}

\input{galw1}
\input{galw2}
\input{galw3}
\input{galw4}
\input{galw5}

\begin{figure}[H] 
\centering
\begin{subfigure}[b]{0.19\textwidth}
    \centering
    \includegraphics[width=\textwidth, trim = {0.5cm 12.5cm 38cm 4.5cm}, clip]{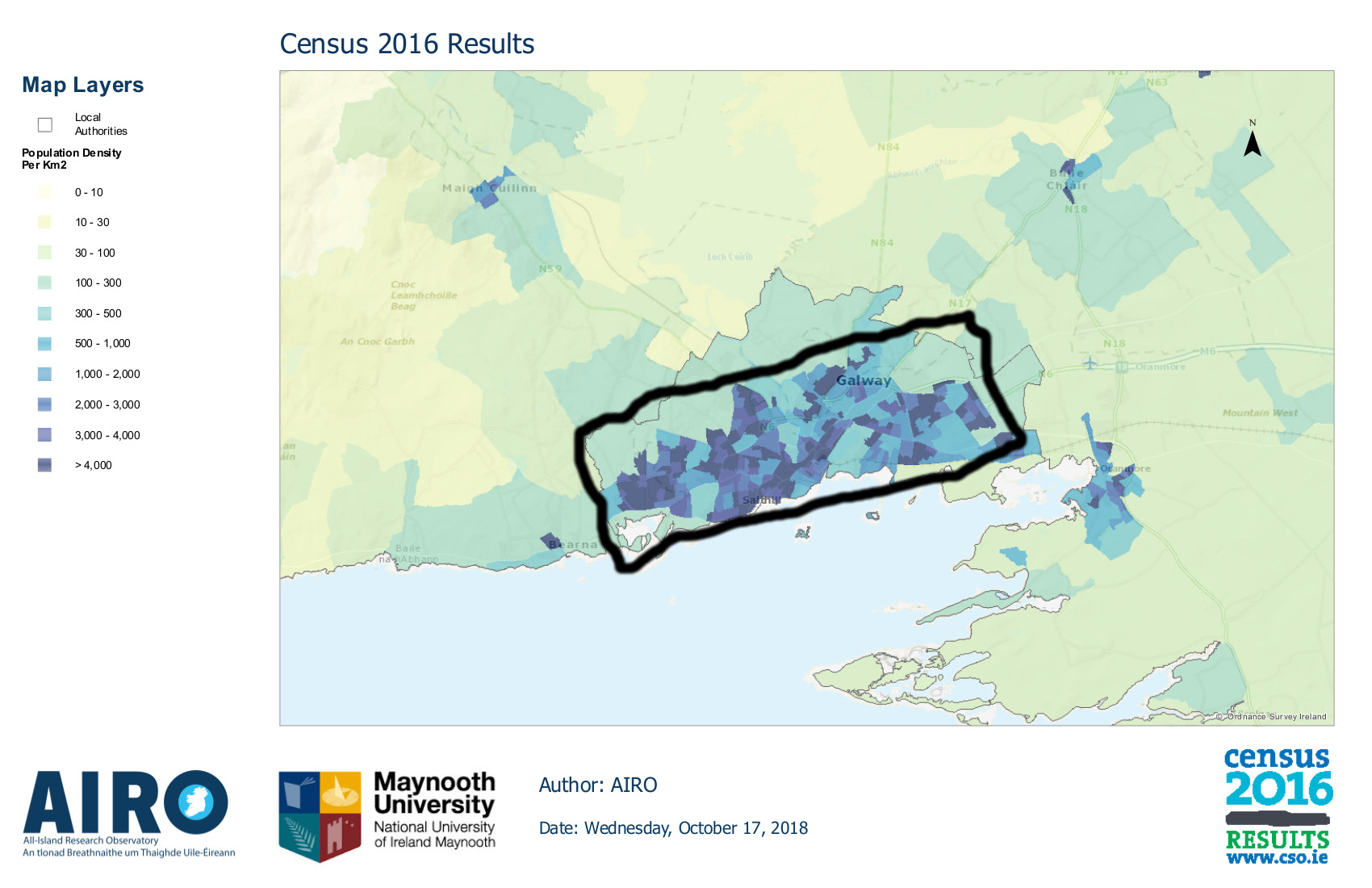}
\end{subfigure}
\begin{subfigure}[b]{0.79\textwidth}  
    \centering 
    \includegraphics[width=\textwidth, trim = {5cm 5cm 0 0}, clip]{galway3}
\end{subfigure}
    \caption[]%
    {{\small Fine-grained population density in Galway City and surroundings. This is taken from \protect\cite{airo, cso1},
    with a superimposed rectangle hand-drawn in black by the author to illustrate the city shape.}}
    \label{blueMap}
\end{figure}
\input{test1}

%% file: map1.tex
\begin{figure}
\centering
\begin{tikzpicture}[x=0.5cm, y=0.5cm]
    \begin{axis}[%
    xmin=10,
    xmax=50,
    ymin=0,
    ymax=0.4,
    hide axis,
    legend style={at={(2.1,0.4)},anchor=north east}
    ]
    \addlegendimage{legend image code/.code={\node [draw, regular polygon, regular polygon sides=4, fill=dens0, minimum size=1em] {};}}
    \addlegendentry{ $> 3500$ per square km};
    \addlegendimage{legend image code/.code={\node [draw, regular polygon, regular polygon sides=4, fill=dens1, minimum size=1em] {};}}
    \addlegendentry{ between 3000 and 3500};
      \addlegendimage{legend image code/.code={\node [draw, regular polygon, regular polygon sides=4, fill=dens2, minimum size=1em] {};}}
    \addlegendentry{ between 2000 and 3000}; 
          \addlegendimage{legend image code/.code={\node [draw, regular polygon, regular polygon sides=4, fill=dens3, minimum size=1em] {};}}
    \addlegendentry{ between 1000 and 2000};  
           \addlegendimage{legend image code/.code={\node [draw, regular polygon, regular polygon sides=4, fill=dens4, minimum size=1em] {};}}
    \addlegendentry{ between 500 and 1000};

    \end{axis}
\node at (-0.5,0.5) { A};
\node at (-0.5,1.5) { B};
\node at (-0.5,2.5) { C};
\node at (-0.5,3.5) { D};
\node at (-0.5,4.5) { E};
\node at (-0.5,5.5) { F};
\node at (-0.5,6.5) { G};
\node at (-0.5,7.5) { H};
\node at (-0.5,8.5) { I};
%
\node at (0.5,9.5) { 1};
\node at (1.5,9.5) {2};
\node at (2.5,9.5) {3};
\node at (3.5,9.5) {4};
\node at (4.5,9.5) {5};
\node at (5.5,9.5) {6};
\node at (6.5,9.5) {7};
\node at (7.5,9.5) {8};
\node at (8.5,9.5) {9};
\node at (9.5,9.5) {10};
\node at (10.5,9.5) {11};
\node at (11.5,9.5) {12};
\node at (12.5,9.5) {13};
\node at (13.5,9.5) {14};
\node at (14.5,9.5) {15};
\node at (15.5,9.5) {16};
\node at (16.5,9.5) {17};
\node at (17.5,9.5) {18};
\node at (18.5,9.5) {19};
\node at (19.5,9.5) {20};

\draw[step=0.5cm, line width=0.5mm, black!50!white] (0,0) grid (\width,\hauteur);
\fill[dens0] (4,5) rectangle (7,6);
\fill[dens0] (8,4) rectangle (9,5);
\fill[dens1] (7,4) rectangle (8,6);
\fill[dens1] (9,6) rectangle (10,7);
\fill[dens1] (12,5) rectangle (13,6);
\fill[dens1] (13,4) rectangle (14,5);
\fill[dens2] (4,4) rectangle (5,5);
\fill[dens2] (6,4) rectangle (7,5);
\fill[dens2] (7,6) rectangle (8,7);
\fill[dens2] (10,6) rectangle (11,7);
\fill[dens2] (9,5) rectangle (10,6);
\fill[dens2] (11,4) rectangle (12,6);
\fill[dens3] (1,4) rectangle (2,5);
\fill[dens3] (3,5) rectangle (4,6);
\fill[dens3] (5,4) rectangle (6,5);
\fill[dens3] (8,5) rectangle (9,6);
\fill[dens3] (10,4) rectangle (11,6);
\fill[dens3] (12,4) rectangle (13,5);
\fill[dens3] (13,5) rectangle (14,6);
\fill[dens3] (14,3) rectangle (15,4);
\fill[dens3] (16,1) rectangle (17,3);
\fill[dens4] (16,0) rectangle (17,1);
\fill[dens4] (15,1) rectangle (16,2);
\fill[dens4] (16,3) rectangle (17,4);
\fill[dens4] (13,3) rectangle (14,4);
\fill[dens4] (11,7) rectangle (12,8);
\fill[dens4] (9,4) rectangle (10,5);
\fill[dens4] (8,6) rectangle (9,7);
\fill[dens4] (7,7) rectangle (8,8);
\fill[dens4] (5,6) rectangle (7,7);
\fill[dens4] (6,3) rectangle (7,4);
\fill[dens4] (3,4) rectangle (4,5);

\node[circle, inner sep=0pt, minimum size=0.3cm, color=black, fill=white] at (11.5,4.5){{\bf a}}; 
\node[circle, inner sep=0pt, minimum size=0.3cm, color=black, fill=white] at (13.5,4.5){{\bf b}}; 
\node[circle, inner sep=0pt, minimum size=0.3cm, color=black, fill=white] at (13.5,5.5){{\bf h}}; 
\node[circle, inner sep=0pt, minimum size=0.3cm, color=black, fill=white] at (5.5,5.5){{\bf c}}; 
\node[circle, inner sep=0pt, minimum size=0.3cm, color=black, fill=white] at (9.5,4.5){{\bf d}}; 
\node[circle, inner sep=0pt, minimum size=0.3cm, color=black, fill=white] at (1.5,4.5){{\bf e}}; 
\node[circle, inner sep=0pt, minimum size=0.3cm, color=black, fill=white] at (16.5,2.5){{\bf f}}; 
\node[circle, inner sep=0pt, minimum size=0.3cm, color=black, fill=white] at (7.5,5.5){{\bf g}}; 
\node [draw,fill=white] at (rel axis cs: 1.85,0.7) {\shortstack[c]{
POINTS OF INTEREST\\
 {\bf a}:GMIT (\emph{Education})\ \ 
 {\bf b}:Galway Clinic (\emph{Hospital})\\
 {\bf c}:Gateway Retail Park (\emph{Retail})\\
 {\bf d}:Main train/bus stations (\emph{Transport})\\
  {\bf e}:Bearna (\emph{Village})\ \ 
 {\bf f}:Oranmore (\emph{Town})\\
  {\bf g}:Westside (\emph{Retail})\ \ 
    {\bf h}:Briarhill (\emph{Retail})
}};

\end{tikzpicture}
 \caption{\small Schematic (drawn to scale) of population densities in Galway City. The figures are taken from
 \protect\cite{survey1}. Grid squares are square kilometres.}
 \label{overGal}
\end{figure}

%% file: data1.tex
\begin{figure}\label{dataGal}
\centering
\begin{tikzpicture}[x=0.5cm, y=0.5cm]
\draw[step=0.5cm, line width=0.3mm, black!90!white] (0,0) grid (\width,\hauteur + 1);
\node at (17.5,0.5) {\NFM{10}};
\node at (16.5,0.5) {\NFM{66}};
\node at (15.5,0.5) {\NFM{5}};
\node at (14.5,0.5) {\NFM{14}};
\node at (13.5,0.5) {\NFM{15}};
\node at (17.5,1.5) {\NFM{6}};
\node at (16.5,1.5) {\NFM{167}};
\node at (15.5,1.5) {\NFM{52}};
\node at (13.5,1.5) {\NFM{2}};
\node at (17.5,2.5) {\NFM{48}};
\node at (16.5,2.5) {\NFM{107}};
\node at (13.5,2.5) {\NFM{1}};
\node at (12.5,2.5) {\NFM{2}};
%
\node at (17.5,3.5) {\NFM{3}};
\node at (16.5,3.5) {\NFM{55}};
\node at (15.5,3.5) {\NFM{11}};
\node at (14.5,3.5) {\NFM{142}};
\node at (13.5,3.5) {\NFM{77}};
\node at (12.5,3.5) {\NFM{21}};
\node at (11.5,3.5) {\NFM{41}};
\node at (8.5,3.5) {\NFM{16}};
\node at (7.5,3.5) {\NFM{24}};
\node at (6.5,3.5) {\NFM{98}};
\node at (5.5,3.5) {\NFM{26}};
\node at (1.5,3.5) {\NFM{3}};
\node at (17.5,4.5) {\NFM{9}};
\node at (16.5,4.5) {\NFM{3}};
\node at (15.5,4.5) {\NFM{8}};
\node at (14.5,4.5) {\NFM{22}};
\node at (13.5,4.5) {\NFM{322}};
\node at (12.5,4.5) {\NFM{152}};
\node at (11.5,4.5) {\NFM{270}};
\node at (10.5,4.5) {\NFM{189}};
\node at (9.5,4.5) {\NFM{59}};
\node at (8.5,4.5) {\NFM{375}};
\node at (7.5,4.5) {\NFM{328}};
\node at (6.5,4.5) {\NFM{256}};
\node at (5.5,4.5) {\NFM{137}};
\node at (4.5,4.5) {\NFM{221}};
\node at (3.5,4.5) {\NFM{90}};
\node at (2.5,4.5) {\NFM{24}};
\node at (1.5,4.5) {\NFM{114}};
\node at (0.5,4.5) {\NFM{40}};
%
\node at (17.5,5.5) {\NFM{4}};
\node at (15.5,5.5) {\NFM{2}};
\node at (14.5,5.5) {\NFM{20}};
\node at (13.5,5.5) {\NFM{133}};
\node at (12.5,5.5) {\NFM{314}};
\node at (11.5,5.5) {\NFM{251}};
\node at (10.5,5.5) {\NFM{139}};
\node at (9.5,5.5) {\NFM{297}};
\node at (8.5,5.5) {\NFM{125}};
\node at (7.5,5.5) {\NFM{323}};
\node at (6.5,5.5) {\NFM{415}};
\node at (5.5,5.5) {\NFM{354}};
\node at (4.5,5.5) {\NFM{397}};
\node at (3.5,5.5) {\NFM{123}};
\node at (2.5,5.5) {\NFM{16}};
\node at (1.5,5.5) {\NFM{12}};
\node at (0.5,5.5) {\NFM{22}};
%
\node at (17.5,6.5) {\NFM{20}};
\node at (16.5,6.5) {\NFM{8}};
\node at (15.5,6.5) {\NFM{6}};
\node at (14.5,6.5) {\NFM{10}};
\node at (13.5,6.5) {\NFM{8}};
\node at (12.5,6.5) {\NFM{5}};
\node at (11.5,6.5) {\NFM{37}};
\node at (10.5,6.5) {\NFM{258}};
\node at (9.5,6.5) {\NFM{306}};
\node at (8.5,6.5) {\NFM{59}};
\node at (7.5,6.5) {\NFM{258}};
\node at (6.5,6.5) {\NFM{85}};
\node at (5.5,6.5) {\NFM{61}};
\node at (4.5,6.5) {\NFM{34}};
\node at (3.5,6.5) {\NFM{9}};
\node at (2.5,6.5) {\NFM{12}};
\node at (1.5,6.5) {\NFM{3}};
\node at (0.5,6.5) {\NFM{3}};
%
\node at (17.5,7.5) {\NFM{9}};
\node at (16.5,7.5) {\NFM{3}};
\node at (15.5,7.5) {\NFM{1}};
\node at (14.5,7.5) {\NFM{13}};
\node at (13.5,7.5) {\NFM{10}};
\node at (12.5,7.5) {\NFM{3}};
\node at (11.5,7.5) {\NFM{69}};
\node at (10.5,7.5) {\NFM{47}};
\node at (9.5,7.5) {\NFM{20}};
\node at (8.5,7.5) {\NFM{3}};
\node at (7.5,7.5) {\NFM{65}};
\node at (6.5,7.5) {\NFM{15}};
\node at (5.5,7.5) {\NFM{11}};
\node at (4.5,7.5) {\NFM{5}};
\node at (3.5,7.5) {\NFM{22}};
\node at (2.5,7.5) {\NFM{5}};
\node at (1.5,7.5) {\NFM{3}};
%
\node at (17.5,8.5) {\NFM{8}};
\node at (16.5,8.5) {\NFM{13}};
\node at (15.5,8.5) {\NFM{14}};
\node at (14.5,8.5) {\NFM{11}};
\node at (13.5,8.5) {\NFM{12}};
\node at (12.5,8.5) {\NFM{6}};
\node at (11.5,8.5) {\NFM{18}};
\node at (9.5,8.5) {\NFM{12}};
\node at (8.5,8.5) {\NFM{32}};
\node at (7.5,8.5) {\NFM{1}};
\node at (6.5,8.5) {\NFM{16}};
\node at (5.5,8.5) {\NFM{11}};
\node at (4.5,8.5) {\NFM{14}};
\node at (3.5,8.5) {\NFM{5}};
\node at (2.5,8.5) {\NFM{3}};
%
\node at (17.5,9.5) {\NFM{40}};
\node at (16.5,9.5) {\NFM{13}};
\node at (15.5,9.5) {\NFM{9}};
\node at (14.5,9.5) {\NFM{5}};
\node at (12.5,9.5) {\NFM{3}};
\node at (11.5,9.5) {\NFM{9}};
\node at (10.5,9.5) {\NFM{24}};
\node at (9.5,9.5) {\NFM{3}};
\node at (6.5,9.5) {\NFM{2}};
\node at (5.5,9.5) {\NFM{13}};
\node at (4.5,9.5) {\NFM{11}};
\node at (3.5,9.5) {\NFM{9}};
\node at (2.5,9.5) {\NFM{15}};
\end{tikzpicture}
 \caption{\small Table / grid of population densities in Galway City. Each square represents a square kilometre, and is drawn to scale. Multiply each number by 10 to get the population in that square.
 (Note that the blank squares at the bottom/left 
 correspond to Galway Bay / Atlantic Ocean - no
 population!)}
 \label{map1}
\end{figure}

%% file: galw1.tex
\begin{figure}[H]
\centering
\begin{tikzpicture}[x=0.5cm, y=0.5cm]
    \begin{axis}[%
    xmin=10,
    xmax=50,
    ymin=0,
    ymax=0.4,
    hide axis,
    legend style={at={(2.1,0.5)},anchor=north east}
    ]
    \addlegendimage{legend image code/.code={\node [draw, regular polygon, regular polygon sides=4, fill=dens0, minimum size=1em] {};}}
    \addlegendentry{ $> 3500$ per square km};
    \end{axis}
\input{nodes.tex}

\draw[step=0.5cm, line width=0.5mm, black!50!white] (0,0) grid (\width,\hauteur);
\fill[dens0] (4,5) rectangle (7,6);
\fill[dens0] (8,4) rectangle (9,5);
\draw[line width = 0.5mm, decorate with=circle, paint=black, decoration={shape evenly spread=7}](4.5,5.5) to [curve through ={(5.5,5.5) . . (6.5,5.5)}] (8.5,4.5);
\draw[line width = 0.5mm](4.5,5.5) to [curve through ={(5.5,5.5) . . (6.5,5.5)}] (8.5,4.5);
\end{tikzpicture}
 \caption{\small Schematic (drawn to scale) of a light rail/tram line connecting only
 the highest density areas in Galway City (see also Figure~\ref{map1}). We consider here
 only areas with more than 3500 people per square kilometre.
 Grid squares are square kilometres. This line services about 19 thousand people (the number of people 
 living within $d_t = 500$ metres of the line).}
 \label{app1}
\end{figure}

%% file: galw2.tex
\begin{figure}[H]
\centering
\begin{tikzpicture}[x=0.5cm, y=0.5cm]
    \begin{axis}[%
    xmin=10,
    xmax=50,
    ymin=0,
    ymax=0.4,
    hide axis,
    legend style={at={(2.1,0.5)},anchor=north east}
    ]
    \addlegendimage{legend image code/.code={\node [draw, regular polygon, regular polygon sides=4, fill=dens0, minimum size=1em] {};}}
    \addlegendentry{ $> 3500$ per square km};
    \addlegendimage{legend image code/.code={\node [draw, regular polygon, regular polygon sides=4, fill=dens1, minimum size=1em] {};}}
    \addlegendentry{ between 3000 and 3500};
    
    \end{axis}
\input{nodes.tex}

\draw[step=0.5cm, line width=0.5mm, black!50!white] (0,0) grid (\width,\hauteur);
\fill[dens0] (4,5) rectangle (7,6);
\fill[dens0] (8,4) rectangle (9,5);
\fill[dens1] (7,4) rectangle (8,6);
\fill[dens1] (9,6) rectangle (10,7);
\fill[dens1] (12,5) rectangle (13,6);
\fill[dens1] (13,4) rectangle (14,5);
\draw[line width = 0.5mm, decorate with=circle, paint=black, decoration={shape evenly spread=18}](4.5,5.5) to 
[curve through ={(5.5,5.5)..(6.5,5.5)..(7.5,5.5)..(7.5,4.5)..(8.5,4.5)..(9.5,6.5)..(12.5,5.5)}] (13.5,4.5);
\draw[line width = 0.5mm](4.5,5.5) to
[curve through ={(5.5,5.5)..(6.5,5.5)..(7.5,5.5)..(7.5,4.5)..(8.5,4.5)..(9.5,6.5)..(12.5,5.5)}] (13.5,4.5);
\end{tikzpicture}
 \caption{\small Schematic (drawn to scale) of a light rail/tram line in Galway City
 connecting 
 areas with more than 3000 people per square kilometre.
 (see also Figure~\ref{map1}). 
 Grid squares are square kilometres.  This line services about 39 thousand people (the number of people 
 living within $d_t = 500$ metres of the line). }
 \label{app3}
\end{figure}

%% file: galw3.tex
\begin{figure}[H]
\centering
\begin{tikzpicture}[x=0.5cm, y=0.5cm]
    \begin{axis}[%
    xmin=10,
    xmax=50,
    ymin=0,
    ymax=0.4,
    hide axis,
    legend style={at={(2.1,0.5)},anchor=north east}
    ]
    \addlegendimage{legend image code/.code={\node [draw, regular polygon, regular polygon sides=4, fill=dens0, minimum size=1em] {};}}
    \addlegendentry{ $> 3500$ per square km};
    \addlegendimage{legend image code/.code={\node [draw, regular polygon, regular polygon sides=4, fill=dens1, minimum size=1em] {};}}
    \addlegendentry{ between 3000 and 3500};
    \addlegendimage{legend image code/.code={\node [draw, regular polygon, regular polygon sides=4, fill=dens2, minimum size=1em] {};}}
    \addlegendentry{ between 2000 and 3000}; 
    
    \end{axis}
\input{nodes.tex}

\draw[step=0.5cm, line width=0.5mm, black!50!white] (0,0) grid (\width,\hauteur);
\fill[dens0] (4,5) rectangle (7,6);
\fill[dens0] (8,4) rectangle (9,5);
\fill[dens1] (7,4) rectangle (8,6);
\fill[dens1] (9,6) rectangle (10,7);
\fill[dens1] (12,5) rectangle (13,6);
\fill[dens1] (13,4) rectangle (14,5);
\fill[dens2] (4,4) rectangle (5,5);
\fill[dens2] (6,4) rectangle (7,5);
\fill[dens2] (7,6) rectangle (8,7);
\fill[dens2] (10,6) rectangle (11,7);
\fill[dens2] (9,5) rectangle (10,6);
\fill[dens2] (11,4) rectangle (12,6);
\draw[line width = 0.5mm, decorate with=circle, paint=black, decoration={shape evenly spread=20}](4.5,4.5) to 
[curve through ={(4.5,5.5)..(5.5,5.5)..(6.5,5.5)..(7.5,5.5)..(7.5,4.5)..(8.5,4.5)..(9.5,6.5)..(10.5,6.5)..(11.5,5.5)..(12.5,5.5)}] (13.5,4.5);
\draw[line width = 0.5mm](4.5,4.5) to
[curve through ={(4.5,5.5)..(5.5,5.5)..(6.5,5.5)..(7.5,5.5)..(7.5,4.5)..(8.5,4.5)..(9.5,6.5)..(10.5,6.5)..(11.5,5.5)..(12.5,5.5)}] (13.5,4.5);
\end{tikzpicture}
 \caption{\small Schematic (drawn to scale) of a light rail/tram line in Galway City 
 connecting 
 areas with more than 2000 people per square kilometre.
 (see also Figure~\ref{map1}). 
 Grid squares are square kilometres. This line services about 42 thousand people (the number of people 
 living within $d_t = 500$ metres of the line). }
 \label{app2}
\end{figure}

%% file: galw4.tex
g\begin{figure}[H]
\centering
\begin{tikzpicture}[x=0.5cm, y=0.5cm]
    \begin{axis}[%
    xmin=10,
    xmax=50,
    ymin=0,
    ymax=0.4,
    hide axis,
    legend style={at={(2.1,0.5)},anchor=north east}
    ]
    \addlegendimage{legend image code/.code={\node [draw, regular polygon, regular polygon sides=4, fill=dens0, minimum size=1em] {};}}
    \addlegendentry{ $> 3500$ per square km};
    \addlegendimage{legend image code/.code={\node [draw, regular polygon, regular polygon sides=4, fill=dens1, minimum size=1em] {};}}
    \addlegendentry{ between 3000 and 3500};
    \addlegendimage{legend image code/.code={\node [draw, regular polygon, regular polygon sides=4, fill=dens2, minimum size=1em] {};}}
    \addlegendentry{ between 2000 and 3000}; 
          \addlegendimage{legend image code/.code={\node [draw, regular polygon, regular polygon sides=4, fill=dens3, minimum size=1em] {};}}
    \addlegendentry{ between 1000 and 2000};  
    
    \end{axis}
\input{nodes.tex}

\draw[step=0.5cm, line width=0.5mm, black!50!white] (0,0) grid (\width,\hauteur);
\fill[dens0] (4,5) rectangle (7,6);
\fill[dens0] (8,4) rectangle (9,5);
\fill[dens1] (7,4) rectangle (8,6);
\fill[dens1] (9,6) rectangle (10,7);
\fill[dens1] (12,5) rectangle (13,6);
\fill[dens1] (13,4) rectangle (14,5);
\fill[dens2] (4,4) rectangle (5,5);
\fill[dens2] (6,4) rectangle (7,5);
\fill[dens2] (7,6) rectangle (8,7);
\fill[dens2] (10,6) rectangle (11,7);
\fill[dens2] (9,5) rectangle (10,6);
\fill[dens2] (11,4) rectangle (12,6);
\fill[dens3] (1,4) rectangle (2,5);
\fill[dens3] (3,5) rectangle (4,6);
\fill[dens3] (5,4) rectangle (6,5);
\fill[dens3] (8,5) rectangle (9,6);
\fill[dens3] (10,4) rectangle (11,6);
\fill[dens3] (12,4) rectangle (13,5);
\fill[dens3] (13,5) rectangle (14,6);
\fill[dens3] (14,3) rectangle (15,4);
\fill[dens3] (16,1) rectangle (17,3);
\draw[line width = 0.5mm, decorate with=circle, paint=black, decoration={shape evenly spread=26}](1.5,4.5) to 
[curve through ={(3.5,5.5)..(4.5,5.5)..(5.5,5.5)..(6.5,5.5)..(7.5,5.5)..(7.5,4.5)..(8.5,4.5)..(9.5,6.5)..(10.5,6.5)..(11.5,5.5)..(12.5,5.5)..(13.5,4.5)..(14.5,3.5)..(16.5,2.5)}] (16.5,1.5);
\draw[line width = 0.5mm](1.5,4.5) to
[curve through ={(3.5,5.5)..(4.5,5.5)..(5.5,5.5)..(6.5,5.5)..(7.5,5.5)..(7.5,4.5)..(8.5,4.5)..(9.5,6.5)..(10.5,6.5)..(11.5,5.5)..(12.5,5.5)..(13.5,4.5)..(14.5,3.5)..(16.5,2.5)}] (16.5,1.5);
\end{tikzpicture}
 \caption{\small Schematic (drawn to scale) of a light rail/tram line in Galway City
 connecting 
 areas with more than 1000 people per square kilometre.
 (see also Figure~\ref{map1}). 
 Grid squares are square kilometres. This line services about 47 thousand people (the number of people 
 living within $d_t = 500$ metres of the line). }
 \label{app4}
\end{figure}

%% file: galw5.tex
\begin{figure}[H]
\centering
\begin{tikzpicture}[x=0.5cm,y=0.5cm]
    \begin{axis}[%
    xmin=10,
    xmax=50,
    ymin=0,
    ymax=0.4,
    hide axis,
    legend style={at={(2.1,0.5)},anchor=north east}
    ]
    \addlegendimage{legend image code/.code={\node [draw, regular polygon, regular polygon sides=4, fill=dens0, minimum size=1em] {};}}
  Recompile
15

    \addlegendentry{ $> 3500$ per square km};
    \addlegendimage{legend image code/.code={\node [draw, regular polygon, regular polygon sides=4, fill=dens1, minimum size=1em] {};}}
    \addlegendentry{ between 3000 and 3500};
    \addlegendimage{legend image code/.code={\node [draw, regular polygon, regular polygon sides=4, fill=dens2, minimum size=1em] {};}}
    \addlegendentry{ between 2000 and 3000}; 
          \addlegendimage{legend image code/.code={\node [draw, regular polygon, regular polygon sides=4, fill=dens3, minimum size=1em] {};}}
    \addlegendentry{ between 1000 and 2000};  
           \addlegendimage{legend image code/.code={\node [draw, regular polygon, regular polygon sides=4, fill=dens4, minimum size=1em] {};}}
    \addlegendentry{ between 500 and 1000};     
    
    \end{axis}
\input{nodes.tex}

\draw[step=0.5cm, line width=0.5mm, black!50!white] (0,0) grid (\width,\hauteur);
\fill[dens0] (4,5) rectangle (7,6);
\fill[dens0] (8,4) rectangle (9,5);
\fill[dens1] (7,4) rectangle (8,6);
\fill[dens1] (9,6) rectangle (10,7);
\fill[dens1] (12,5) rectangle (13,6);
\fill[dens1] (13,4) rectangle (14,5);
\fill[dens2] (4,4) rectangle (5,5);
\fill[dens2] (6,4) rectangle (7,5);
\fill[dens2] (7,6) rectangle (8,7);
\fill[dens2] (10,6) rectangle (11,7);
\fill[dens2] (9,5) rectangle (10,6);
\fill[dens2] (11,4) rectangle (12,6);
\fill[dens3] (1,4) rectangle (2,5);
\fill[dens3] (3,5) rectangle (4,6);
\fill[dens3] (5,4) rectangle (6,5);
\fill[dens3] (8,5) rectangle (9,6);
\fill[dens3] (10,4) rectangle (11,6);
\fill[dens3] (12,4) rectangle (13,5);
\fill[dens3] (13,5) rectangle (14,6);
\fill[dens3] (14,3) rectangle (15,4);
\fill[dens3] (16,1) rectangle (17,3);
\fill[dens4] (16,0) rectangle (17,1);
\fill[dens4] (15,1) rectangle (16,2);
\fill[dens4] (16,3) rectangle (17,4);
\fill[dens4] (13,3) rectangle (14,4);
\fill[dens4] (11,7) rectangle (12,8);
\fill[dens4] (9,4) rectangle (10,5);
\fill[dens4] (8,6) rectangle (9,7);
\fill[dens4] (7,7) rectangle (8,8);
\fill[dens4] (5,6) rectangle (7,7);
\fill[dens4] (6,3) rectangle (7,4);
\fill[dens4] (3,4) rectangle (4,5);
\draw[line width = 0.5mm, decorate with=circle, paint=black, decoration={shape evenly spread=26}](1.5,4.5) to 
[curve through ={(3.5,4.5)..(4.5,5.5)..(5.5,5.5)..(6.5,5.5)..(7.5,5.5)..(7.5,4.5)..(8.5,4.5)..(9.5,6.5)..(10.5,6.5)..(11.5,5.5)..(12.5,5.5)..(13.5,4.5)..(14.5,3.5)..(16.5,2.5)}] (16.5,1.5);
\draw[line width = 0.5mm](1.5,4.5) to
[curve through ={(3.5,4.5)..(4.5,5.5)..(5.5,5.5)..(6.5,5.5)..(7.5,5.5)..(7.5,4.5)..(8.5,4.5)..(9.5,6.5)..(10.5,6.5)..(11.5,5.5)..(12.5,5.5)..(13.5,4.5)..(14.5,3.5)..(16.5,2.5)}] (16.5,1.5);
\end{tikzpicture}
 \caption{\small Schematic (drawn to scale) of a light rail/tram line in Galway City
 connecting 
 areas with more than 500 people per square kilometre.
 (see also Figure~\ref{map1}). 
 Grid squares are square kilometres. 
 This line services about 48 thousand people (the number of people 
 living within $d_t = 500$ metres of the line).
 }
 \label{app5}
\end{figure}

%% file: test1.tex
\begin{figure}[hbt!]
\centering
\scalebox{1.2}
{
\begin{tikzpicture}[x=0.6cm,y=0.6cm]
    \node[anchor=south west,inner sep=0] (image) at (-2.3,-2.5) {\includegraphics[clip, trim=6cm 17cm 0.1cm 3cm, width=0.75\textwidth, angle=-16]{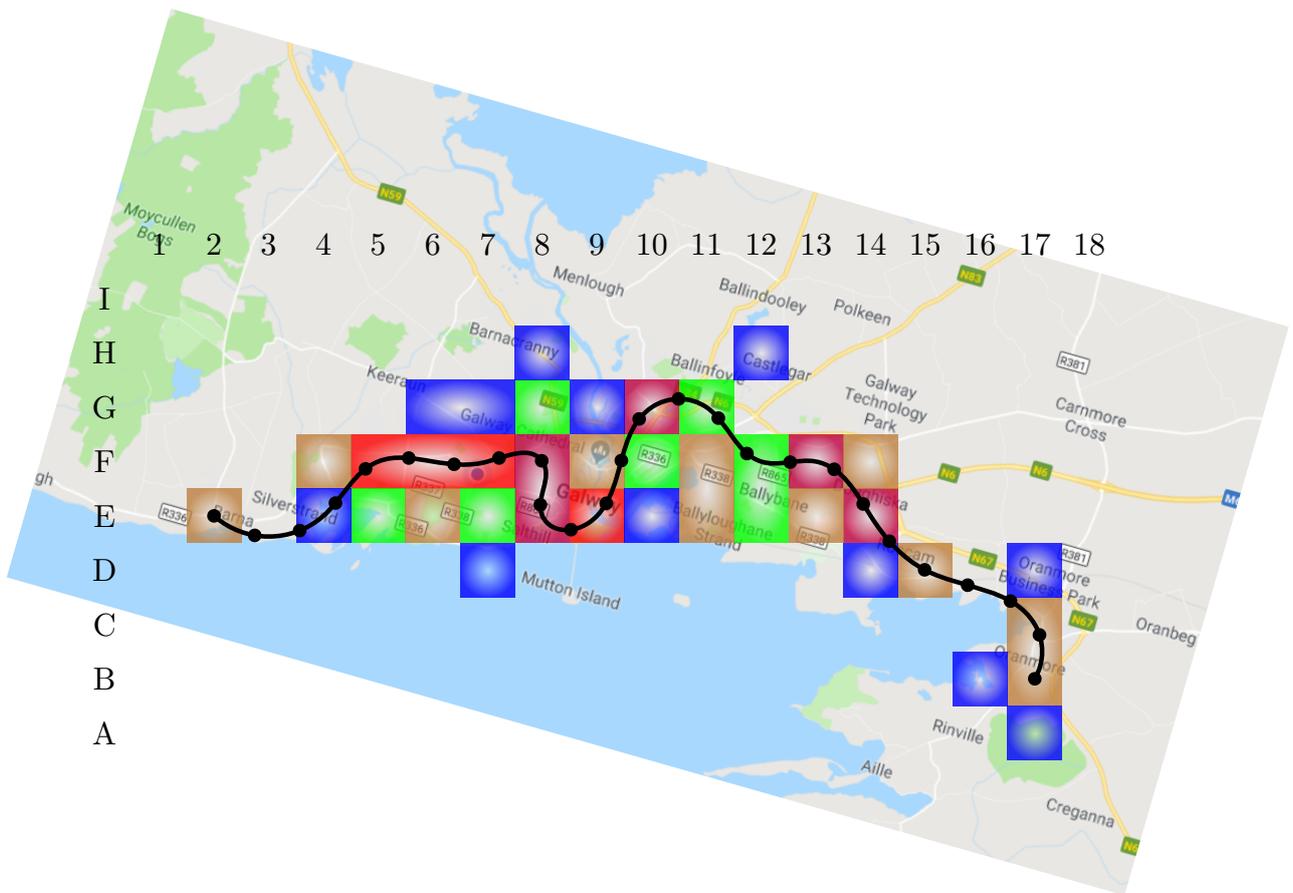}};
    \begin{axis}[%
    width=0.4\textwidth,
    height=0.4\textwidth,
    hide axis,
    legend style={at={(2.1,0.5)},anchor=north east}
    ]
    \addlegendimage{legend image code/.code={\node [draw, regular polygon, regular polygon sides=4, fill=dens0, minimum size=1em] {};}}
    \addlegendentry{ $> 3500$ per square km};
    \addlegendimage{legend image code/.code={\node [draw, regular polygon, regular polygon sides=4, fill=dens1, minimum size=1em] {};}}
    \addlegendentry{ between 3000 and 3500};
    \addlegendimage{legend image code/.code={\node [draw, regular polygon, regular polygon sides=4, fill=dens2, minimum size=1em] {};}}
    \addlegendentry{ between 2000 and 3000}; 
          \addlegendimage{legend image code/.code={\node [draw, regular polygon, regular polygon sides=4, fill=dens3, minimum size=1em] {};}}
    \addlegendentry{ between 1000 and 2000};  
           \addlegendimage{legend image code/.code={\node [draw, regular polygon, regular polygon sides=4, fill=dens4, minimum size=1em] {};}}
    \addlegendentry{ between 500 and 1000};     
    
    \end{axis}
\input{nodes.tex}

\fill[dens0,path fading=fadeIn] (4,5) rectangle (7,6);
\fill[dens0,path fading=fadeIn] (8,4) rectangle (9,5);
\fill[dens1,path fading=fadeIn] (7,4) rectangle (8,6);
\fill[dens1,path fading=fadeIn] (9,6) rectangle (10,7);
\fill[dens1,path fading=fadeIn] (12,5) rectangle (13,6);
\fill[dens1,path fading=fadeIn] (13,4) rectangle (14,5);
\fill[dens2,path fading=fadeIn] (4,4) rectangle (5,5);
\fill[dens2,path fading=fadeIn] (6,4) rectangle (7,5);
\fill[dens2,path fading=fadeIn] (7,6) rectangle (8,7);
\fill[dens2,path fading=fadeIn] (10,6) rectangle (11,7);
\fill[dens2,path fading=fadeIn] (9,5) rectangle (10,6);
\fill[dens2,path fading=fadeIn] (11,4) rectangle (12,6);
\fill[dens3,path fading=fadeIn] (1,4) rectangle (2,5);
\fill[dens3,path fading=fadeIn] (3,5) rectangle (4,6);
\fill[dens3,path fading=fadeIn] (5,4) rectangle (6,5);
\fill[dens3,path fading=fadeIn] (8,5) rectangle (9,6);
\fill[dens3,path fading=fadeIn] (10,4) rectangle (11,6);
\fill[dens3,path fading=fadeIn] (12,4) rectangle (13,5);
\fill[dens3,path fading=fadeIn] (13,5) rectangle (14,6);
\fill[dens3,path fading=fadeIn] (14,3) rectangle (15,4);
\fill[dens3,path fading=fadeIn] (16,1) rectangle (17,3);
\fill[dens4,path fading=fadeIn] (16,0) rectangle (17,1);
\fill[dens4,path fading=fadeIn] (15,1) rectangle (16,2);
\fill[dens4,path fading=fadeIn] (16,3) rectangle (17,4);
\fill[dens4,path fading=fadeIn] (13,3) rectangle (14,4);
\fill[dens4,path fading=fadeIn] (11,7) rectangle (12,8);
\fill[dens4,path fading=fadeIn] (9,4) rectangle (10,5);
\fill[dens4,path fading=fadeIn] (8,6) rectangle (9,7);
\fill[dens4,path fading=fadeIn] (7,7) rectangle (8,8);
\fill[dens4,path fading=fadeIn] (5,6) rectangle (7,7);
\fill[dens4,path fading=fadeIn] (6,3) rectangle (7,4);
\fill[dens4,path fading=fadeIn] (3,4) rectangle (4,5);
\draw[line width = 0.5mm, decorate with=circle, paint=black, decoration={shape evenly spread=26}](1.5,4.5) to 
[curve through ={(3.5,4.5)..(4.5,5.5)..(5.5,5.5)..(6.5,5.5)..(7.5,5.5)..(7.5,4.5)..(8.5,4.5)..(9.5,6.5)..(10.5,6.5)..(11.5,5.5)..(12.5,5.5)..(13.5,4.5)..(14.5,3.5)..(16.5,2.5)}] (16.5,1.5);
\draw[line width = 0.5mm](1.5,4.5) to
[curve through ={(3.5,4.5)..(4.5,5.5)..(5.5,5.5)..(6.5,5.5)..(7.5,5.5)..(7.5,4.5)..(8.5,4.5)..(9.5,6.5)..(10.5,6.5)..(11.5,5.5)..(12.5,5.5)..(13.5,4.5)..(14.5,3.5)..(16.5,2.5)}] (16.5,1.5);

\end{tikzpicture}
}
 \caption{Population density grid squares and tram line superimposed on GOOGLE map of Galway City. The legend for the colored kilometre squares (corresponding to population densities) is as
 used previously
 (e.g.\ Figure \ref{app5}).
 }\label{googleG}
\end{figure}